\documentclass[iop]{emulateapj}
\pdfoutput=1
\usepackage{graphicx,amssymb,epsfig}
\usepackage[autostyle]{csquotes}
\usepackage{setspace}
\usepackage{natbib}
\usepackage{placeins}
\usepackage[utf8]{inputenc}

\slugcomment{Accepted for publication in ApJ October 7, 2014}

\shortauthors{Krogager et al.}

\begin{document}

\title{A spectroscopic sample of massive, evolved $\lowercase{z\sim2}$ galaxies: Implications for the evolution of the mass--size relation\altaffilmark{1}}

\author{
J.-K. Krogager\altaffilmark{2,3},
A. W. Zirm\altaffilmark{2},
S. Toft\altaffilmark{2},
A. Man\altaffilmark{2},
G. Brammer\altaffilmark{3,4}
}

\altaffiltext{1}{Based on observations carried out under programs
\#12177, 12328 with the Wide Field Camera 3 installed on the {\it Hubble Space Telescope}.}
\altaffiltext{2}{Dark Cosmology Centre, Niels Bohr Institute, University of
Copenhagen, Juliane Maries Vej 30, DK-2100 Copenhagen O}
\altaffiltext{3}{European Southern Observatory, Alonso de C\'ordova 3107,
Casilla 19001, Vitacura, Santiago, Chile}
\altaffiltext{4}{Space Telescope Science Institute, 3700 San Martin Drive,
Baltimore, MD21210, USA}

\begin{abstract}
We present deep, near-infrared {\it HST}/WFC3 grism spectroscopy and imaging for
a sample of 14 galaxies at $z\approx2$ selected from a mass-complete photometric
catalog in the COSMOS field.
By combining the grism observations with photometry in 30 bands, we derive
accurate constraints on their redshifts, stellar masses, ages, dust
extinction and formation redshifts. 
We show that the slope and scatter of the $z\sim2$ mass--size
relation of quiescent galaxies is consistent with the local relation, and confirm previous
findings that the sizes for a given mass are smaller by a factor of two to three. 
Finally, we show that the observed evolution of the mass--size relation of
quiescent galaxies between $z=2$ and $0$ can be explained by 
quenching of increasingly larger star-forming galaxies, at a rate
dictated by the increase in the number density of quiescent
galaxies with decreasing redshift. However, we find that the scatter in the mass--size
relation should increase in the quenching-driven scenario in contrast to what is seen
in the data. This suggests that merging is not needed to explain the
evolution of the median mass--size relation of massive galaxies, but may
still be required to tighten its scatter, and explain the size growth
of individual $z=2$ galaxies quiescent galaxies.
\end{abstract}

\keywords{galaxies: formation --- galaxies: high-redshift --- cosmology: observations}

\section{Introduction}
Over the past decade, studies of the $z \sim 2$ Universe have been
revolutionized by the availability of deep near-infrared (NIR) imaging
surveys. 
One of the primary early results was the discovery of a population of
optically-faint, massive galaxies which are missed in optical
(rest-frame UV) surveys \citep{Franx2003,Daddi2004a,Wuyts2007}. 
Large photometric surveys have since shown that at $z=2$, roughly half of
the most massive ($\log {\rm M/M_{\odot}}>11$) galaxies are dusty and star-forming,
and half are old, quiescent systems \citep[e.g.][]{Franx2008, Toft2009, Williams2010, Brammer2011},
a result that has been confirmed through low resolution spectroscopy of a small sample of the brightest
examples \citep{Kriek2008a, Kriek2009b, Kriek2009a}. 

Using high-resolution NIR imaging, it was shown that most of
the quiescent galaxies at $z > 2$ have effective radii, $r_e$, a factor of $2-6$ smaller
than local elliptical galaxies with the same stellar masses \citep[e.g.,][]{Daddi2005,
Trujillo2006, Zirm2007, Toft2007, Buitrago2008, vanDokkum2008,
Szomoru2010, Cassata2011, Bruce2012, Newman2012}.
Their inferred stellar mass densities (within $r_e$) therefore greatly exceed those
of local galaxies at the same stellar mass. However, recent studies show
that if one compares the stellar densities within the inner 1~kpc the discrepancy
is much less pronounced \citep{Bezanson2009,Patel2013}. The discovery that the inner regions of these
massive galaxies correspond well with their local counterparts supports
the so-called {\it inside-out} scenario, in which galaxies form at high redshift as
compact galaxies presumably from a gas rich merger funneling the gas to the center
and igniting a massive, compact star burst \citep[e.g.,][]{Hopkins2006, Wuyts2010}.
These resulting compact stellar cores subsequently grow by adding mass to their
outer regions. How this size growth is accomplished is the big question; A cascade of merger
events with smaller systems, known as minor merging, is a plausible explanation
as simulations have shown that it is possible to obtain the needed mass increase
in the outer regions while leaving the central core mostly intact
\citep{Oser2012}. However, observations of the merger rate of massive
galaxies between $z=2$ and $0$ do not find as many mergers as required to
account for the observed size evolution \citep{Man2012, Newman2012}.

Some studies of high-redshift, quenched galaxies have suggested that their
structure may differ from that of local elliptical galaxies when quantified using a S\'ersic
profile. \citet{Chevance2012} find that the high-$z$ galaxies show lower S\'ersic indices ($n\sim2$ on average)
than the local population of ellipticals ($n\sim4$), see also \citet{Weinzirl2011}.
This is further supported by
the detection of galaxies at redshift $z\sim2.5$ with apparently disc-like morphologies \citep{Stockton2004, Stockton2008}.
This has motivated suggestions that the high-$z$ population might be more disc-like and hence
might contain a faint, extended stellar component which would be
undetected in present observations due to cosmological surface-brightness dimming \citep{vanderWel2011}.
Deeper and higher resolution imaging, along with image stacking, has confirmed
that the massive, red galaxies indeed are compact, and has failed to detect any
extended stellar haloes around these compact cores \citep{vanDokkum2008,vanDokkum2010}.
Moreover, \citet{Szomoru2012} find S\'ersic indices for galaxies at high redshift ($z\sim2$)
consistent with the local values. They report a median value for the S\'ersic indices at high redshift of $n=3.7$.

Now, with the advent of the next generation of NIR spectrographs on
8-m class telescopes, we can study the stellar populations via
continuum detections and absorption line indices \citep[][Zirm et al. in prep]{Toft2012,
Onodera12, Sande2013, Belli2014a, Belli2014b}; The
quiescent galaxies can be further sub-divided into post-starbursts
(those that show strong Balmer absorption lines) and more evolved
systems with metal absorption lines. However, even with state-of-the-art
instrumentation, target samples are limited to the rare and bright examples.

Grism spectroscopy from space with {\it Hubble Space Telescope} ({\it HST}) allows us to
obtain redshifts for fainter, less massive examples of $z \sim
2$ galaxies. While these data have poor spectral resolution, they do
not suffer from the strong atmospheric emission lines, poor
transmission and bright background that limit ground-based observations.
A near-infrared spectroscopy survey, 3D-HST, has recently been carried out
using the Wide Field Camera 3 (WFC3) onboard the {\it HST}. The survey
provides imaging in the F140W-band and grism observations in the G141
grism. In total the survey provides rest-frame optical spectra of $\sim$7000 galaxies
in the redshift range from $z=1-3.5$. Moreover, the pointings cover approximately three
quarters of the deep NIR survey, CANDELS \citep{Grogin11, Koekemoer11}. The combination
of imaging and spectroscopic data from 3D-HST and CANDELS allows for 
powerful analysis of the redshift $1<z<3.5$ Universe. For more details
about the 3D-HST survey, see \citet{Brammer12}.

Until now, spectroscopic samples of quiescent, high-redshift galaxies with structural
data have been sparse; \citet{vanDokkum2008} presented a sample of 
nine galaxies at $z\sim2$, \citet{Gobat2013} presented five quiescent galaxies
from a proto-cluster at $z=2$, \citet{Tanaka2013} presented spectroscopy of a $z=2.16$ proto-cluster
with four quiescent members, and recently \citet{Belli2014b} presented a sample of 6 quiescent galaxies
at $2<z<2.5$. At slightly lower redshifts
\citet{Onodera12} presented sample of 18 quiescent galaxies at $z\sim1.6$ and \citet{Belli2014a} presented
a large sample of 103 galaxies with redshifts $0.9<z<1.6$.
Samples of $z\sim2$ quiescent galaxies with measured velocity dispersions
and dynamical masses are even smaller; so far only a handful of examples have
been published \citep[][]{vanDokkum2009, Onodera2010,Sande2011,Toft2012, Belli2014b}.

We have compiled a mass-complete sample ($\log({\rm M/M}_{\odot})>10.8$) of galaxies in
the COSMOS region of the 3D-HST survey. By matching the photometric sample of galaxies
to the spectra extracted from the 3D-HST data we can improve the redshift determinations
compared to the photometric redshifts used so far.
By inferring sizes, redshifts, and stellar population parameters including 
age, star-formation rate, and mass, we are able to populate the mass--size relation
using a mass-complete, quiescent sample of galaxies at $z \sim 2$.
This provides strong constraints on what drives the size evolution of the
massive galaxies.
We explore different physical explanations for the apparent size
growth. Specifically, we create a simplistic model to investigate the
effect of progenitor bias, i.e., addition of newly quenched, larger
galaxies to the mass--size relation, a mechanism proposed by previous
studies \citep{Cassata2011, Trujillo2012, Poggianti2013} and recently
investigated in detail out to redshift $z\sim1$ by \cite{Carollo2013}.

The paper is organized as follows: In section 2 we present the data used in
our analysis, in section 3 we describe the selection of our sample before
presenting the results of our analysis in section 4 and 5, in section 6 we investigate
the mass--size relation and describe our model for size evolution driven by quenching,
and finally we discuss the implications in section 7.

Throughout this paper, we assume a flat cosmology with
$\Omega_{\Lambda}=0.73$, $\Omega_m=0.27$ and a Hubble constant of
$H_0=71$~km~s$^{-1}$~Mpc$^{-1}$.

\section{Data}
\label{data}
The analysis is based on public grism spectroscopy data from the 3D-HST survey
from which we have used 25 pointings in the COSMOS field. We have combined the
spectroscopic data with photometric data in 30 bands covering 0.15--24~$\mu$m
from the latest $K_{\rm s}$-selected catalog by \cite{Muzzin2013}.

The 25 pointings in COSMOS are covered by imaging in the F140W filter and by
NIR spectroscopy using the G141 grism providing wavelength coverage from $1.1~\mu$m
to $1.6~\mu$m with a spectral resolution of $R\sim130$ (for a point source) with
a sampling of $46.5$~{\AA} per pixel.
Since these are slitless spectroscopic data the effective resolution depends on the
size and morphology of the dispersed source. This results in an effective resolution
of the order of $R\sim50$ due to morphological broadening.
Furthermore, we have used the structural parameters from \citet{vanderWel2012}
obtained from the WFC3/F160W ($H_{160}$) images as part of the public CANDELS
survey \citep{Grogin11, Koekemoer11}.

\subsection{Data reduction}
\label{reduction}
Each pointing was observed in a four-point dither pattern with half-pixel offsets
in order to increase the resolution of the final image. Both the undispersed, direct
images and the dispersed grism images were observed with this pattern for a total
exposure of around 800~sec and 4700~sec for undispersed and dispersed, respectively.

The data sets were reduced using the publicly available pipeline {\it threedhst}
\footnote{http://code.google.com/p/threedhst} \citep{Brammer12}. The pipeline handles
the combination and reduction of the dithered exposures, source identification
using {\sc SExtractor}, and extraction of the individual spectra.
Since we are dealing with slitless spectroscopy some sources will have spatially overlapping
spectra. This is handled in the pipeline and each extracted spectrum is provided with an
estimate of the amount of contamination from nearby sources. For our analysis, we have
subtracted the contaminating flux from the total extracted flux.

We have used the standard extraction parameters in the pipeline except for the final pixel scale
used in the call to the {\sc iraf}-task {\sc multidrizzle}, where we chose $0\farcs09$~px$^{-1}$
instead of $0\farcs06$~px$^{-1}$.
This was chosen to reduce the noise in the extracted spectra. For further details about the
observations and data reduction see \citet{Brammer12}. We used a detection threshold of
$4~\sigma$ to identify sources in the F140W images.

After the initial reduction we encountered some issues with the background not being flat. We were
not able to correct this gradient sufficiently to recover a completely flat background, which meant that
some spectra were disregarded due to background issues. However, when we increased the pixel size from
$0\farcs06$~px$^{-1}$ to $0\farcs09$~px$^{-1}$ the noise decreased and the background subtraction
was performed more successfully. In the process of selecting our sample we removed three sources due
to background-subtraction issues. In these cases (IDs \#133915, \#133784, and \#207144)
there were discontinuities in the background, that we
could not correct for. See appendix \ref{sample:notes} for details.

\section{Sample Selection}
\label{sel}
We selected all sources in the survey area (25 pointings in the COSMOS field from the 3D-HST survey)
from a mass complete sample by Muzzin et al. 2013 with stellar masses larger than $\log(\rm{M_{\star}/M_{\odot}}) > 10.8$. Only sources with photometric redshifts in the range of $1.85<z_{\rm phot}<2.3$ were included in order to sort out low-redshift interlopers. The redshift cut ensured that the 4000 {\AA} break will be visible in the spectra, as this is the only strong spectral feature available at this resolution and wavelength coverage. In total we have 34 galaxies above the mass limit with photometric data.

The photometric targets were then matched by coordinates to the catalog of extracted sources in the 3D-HST data. 
Next step was to sort out incomplete spectra. We required that at least 80per cent of the pixels be well-defined, i.e., non-zero and non-negative. In some cases where the objects were located close to the edge of the CCD some light was dispersed
out over the edges, and hence the spectral range was reduced in those cases.
By only allowing spectra with more than 80 per cent well-defined pixels, we ensured that our targets were
fully covered in the wavelength range from 1.1--1.6~$\mu$m. For each spectrum, we calculated the median SNR per pixel. We discarded two targets (IDs \#123780 and \#124168) with a median SNR per pixel less than 1, i.e., more than half of the pixels were heavily noise dominated.

The sources with spectra were then scaled to the $J$-band flux, which is fully covered by $G$141. Sources
whose $J$-band flux could not be calculated due to too many noisy pixels within the $J$-band wavelength coverage were discarded.
We scaled the spectra to account for the loss of flux due to the limitations of the spectral extraction aperture.
Finally, we checked how well the contamination (if any) had been subtracted. We discarded the most
heavily contaminated spectra and the spectra where the contamination had been subtracted incorrectly
leaving gaps and holes in the extracted spectra.
This was done by visual inspection as not only the amount of contamination is important, but also the shape
of the contaminating flux. In some cases the contaminating flux can enhance or even create a break in the
spectrum, and this is difficult to quantify in a comparable way for all targets.

In the end we end up with a spectroscopic sample of 14 galaxies. The properties of the full sample consisting of 34 galaxies (of which 14 have spectral data available) are summarized in Table~\ref{sample}. In appendix \ref{sample:notes}, we give a list of the spectra that were discarded along with remarks on why they were removed from our sample.

\subsection{Quiescent versus Star-Forming}

Since we only wish to study the quiescent population, we need at method to distinguish between star-forming and quiescent galaxies. For this we use the so-called $UVJ$ selection, based on the rest-frame $U-V$ and $V-J$ colours following \citet{Williams2009}. The rest-frame colours for our sample are included in the catalog from \citet{Muzzin2013}. The classification from the $UVJ$ criteria is given for each sample member in Table~\ref{sample}: 'QG' for quiescent galaxies and 'SFG' for star-forming galaxies.

In Sect.~\ref{fast}, we will analyse the spectral energy distributions (SEDs) and compare the $UVJ$-classification with the star formation rates inferred from the SEDs.

\subsection{Spectral completeness}
\begin{figure*}
  \includegraphics[width=0.96\textwidth]{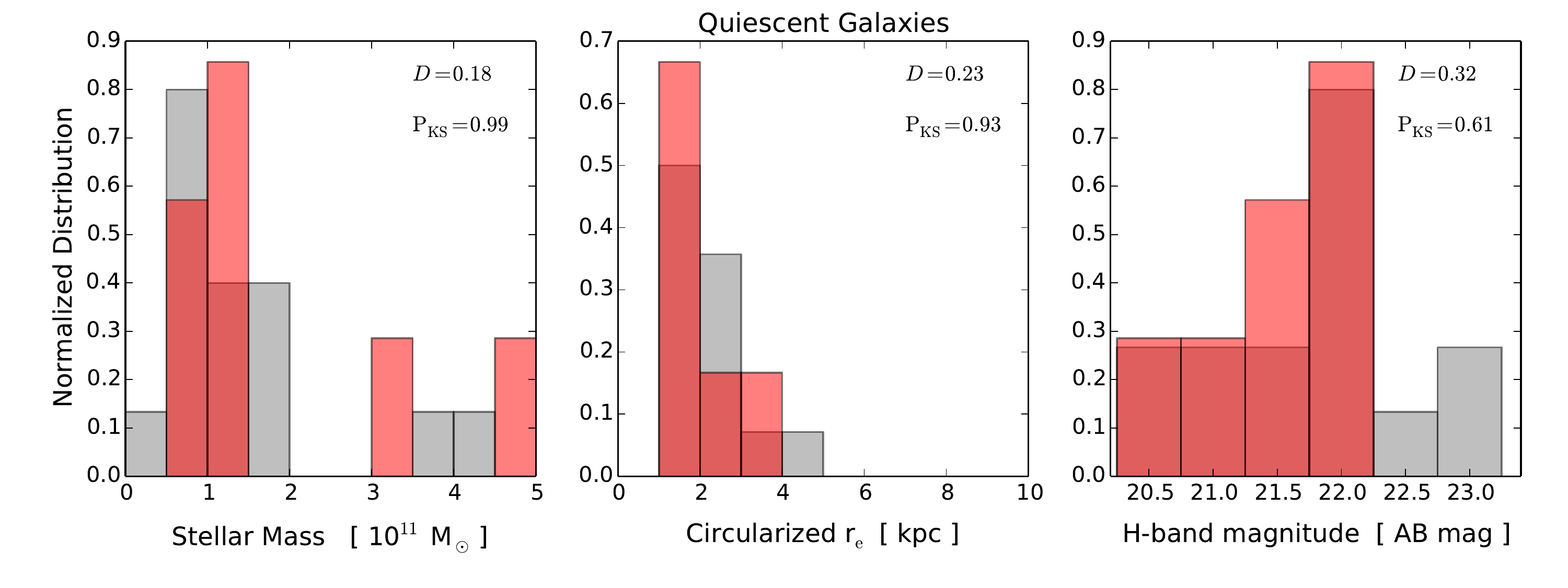}
  \includegraphics[width=0.96\textwidth]{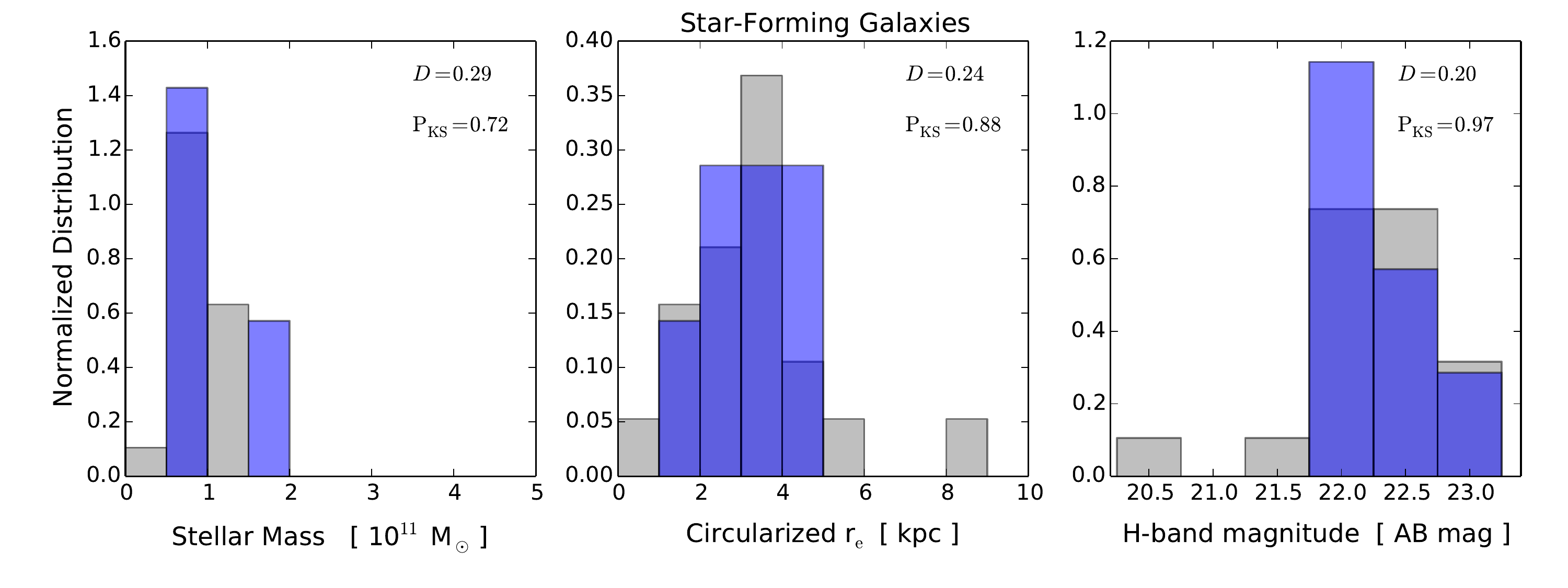}
  \caption{Normalized distributions of stellar mass, effective radii, and $H$-band AB magnitudes. In each panel, the underlying gray histogram shows the full photometric parent sample of quiescent galaxies, which was selected from the mass-complete catalog. The red histogram shows the spectroscopic sub-sample. In the corner of each panel, we show the KS statistic, $D$, and the $p$-value, $P_{\rm KS}$. The top row shows the quiescent galaxies and the lower row shows the star-forming galaxies.}
   \label{fig:complete_SFG}
\end{figure*}

In the process of discarding galaxies from the photometric sample based on their available spectra we may have introduced biases in the final spectroscopic sample, e.g., the sizes and masses may be systematically underestimated due to low SNR and faint sources being discarded. In this section, we address the selection effects. We perform a Kolmogorov-Smirnov (KS) test on the distributions of masses, sizes and $H$-band magnitudes comparing the spectroscopic sub-sample to the full photometric sample, the details as to how we infer masses and sizes are described in Sect.~\ref{fast} and \ref{galfit}.
In Fig.~\ref{fig:complete_SFG}, we show the distributions of stellar mass, circularized effective radii and $H$-band AB magnitudes for quiescent and star-forming galaxies. In each panel, we give the calculated KS statistic along with the $p$ value. All distributions are fully consistent with being drawn from the same distribution, and hence no significant bias is introduced in the spectroscopic selection. Nevertheless, we do note that the spectroscopic selection for the quiescent galaxies seems to exclude the faintest sources with $H>22.25$. However, this does not translate into a bias in the inferred masses and sizes. On the other hand, the brightest star-forming galaxies seem to be discarded by the spectroscopic selection. This may be explained by the star-forming galaxies on average being more extended and thus having their flux distributed among more noisy pixels, thereby lowering the SNR in the spectrum.

\begin{table*}[h]
\caption{Description of the full sample. The galaxy type refers to the UVJ classification as either star-forming (SFG) or quiescent (QG). \label{sample}}
\begin{center}
\setstretch{1.3}
\begin{tabular}{lcccccl}
\hline
\hline
Catalog ID & Grism ID & RA & DEC & z$_{\mathrm{phot}}$ & H$_{160}$ & Type \\ 
 &  & (deg) & (deg) &  & ({\small AB} mag) &  \\ 
\hline
118543 & ibhm42.014 & 150.1039700 & 2.1864245 & 2.09 $^{+0.12}_{-0.12}$ & 22.50 $\pm$ 0.10 & SFG \\ 
119753 &  & 150.1430800 & 2.2003083 & 1.99 $^{+0.17}_{-0.16}$ & 22.00 $\pm$ 0.09 & QG \\ 
121761 & ibhm42.243 & 150.1172600 & 2.2238793 & 1.97 $^{+0.11}_{-0.10}$ & 21.96 $\pm$ 0.05 & SFG \\ 
122398 & ibhm30.211 & 150.1539200 & 2.2323158 & 1.95 $^{+0.10}_{-0.09}$ & 21.84 $\pm$ 0.05 & QG \\ 
123235 & ibhm41.170 & 150.0981300 & 2.2428155 & 2.01 $^{+0.21}_{-0.20}$ & 22.85 $\pm$ 0.13 & SFG \\ 
123324 &  & 150.0905500 & 2.2441640 & 2.26 $^{+0.11}_{-0.12}$ & 22.54 $\pm$ 0.12 & SFG \\ 
123780 &  & 150.0768000 & 2.2492609 & 2.26 $^{+0.20}_{-0.20}$ & 22.42 $\pm$ 0.18 & SFG \\ 
123817 &  & 150.0765800 & 2.2496693 & 2.13 $^{+0.13}_{-0.13}$ & 22.36 $\pm$ 0.11 & SFG \\ 
124168 &  & 150.1064100 & 2.2515926 & 2.04 $^{+0.08}_{-0.10}$ & 22.65 $\pm$ 0.13 & SFG \\ 
124666 & ibhm33.161 & 150.0656400 & 2.2609785 & 1.98 $^{+0.10}_{-0.11}$ & 21.04 $\pm$ 0.04 & QG \\ 
124686 &  & 150.0639300 & 2.2611351 & 2.11 $^{+0.17}_{-0.13}$ & 21.95 $\pm$ 0.06 & QG \\ 
126073 &  & 150.1429400 & 2.2785251 & 2.21 $^{+0.11}_{-0.09}$ & 22.15 $\pm$ 0.07 & QG \\ 
126301 &  & 150.0778700 & 2.2811451 & 2.05 $^{+0.11}_{-0.11}$ & 22.10 $\pm$ 0.07 & SFG \\ 
126824 & ibhm40.143 & 150.1213500 & 2.2851899 & 2.19 $^{+0.13}_{-0.13}$ & 22.06 $\pm$ 0.09 & SFG \\ 
126952 &  & 150.0705900 & 2.2892091 & 2.28 $^{+0.27}_{-0.29}$ & 23.21 $\pm$ 0.22 & SFG \\ 
127466 & ibhm51.200 & 150.1553000 & 2.2948477 & 1.97 $^{+0.14}_{-0.12}$ & 22.01 $\pm$ 0.07 & QG \\ 
127603 & ibhm51.292 & 150.1593200 & 2.2968187 & 1.96 $^{+0.19}_{-0.17}$ & 22.60 $\pm$ 0.11 & SFG \\ 
127617 &  & 150.1196100 & 2.2958610 & 2.07 $^{+0.07}_{-0.07}$ & 20.73 $\pm$ 0.02 & SFG \\ 
128061 & ibhm54.240 & 150.0739400 & 2.2979755 & 2.09 $^{+0.02}_{-0.02}$ & 20.44 $\pm$ 0.02 & QG \\ 
128093 & ibhm54.256 & 150.0746200 & 2.3020012 & 2.18 $^{+0.10}_{-0.09}$ & 21.87 $\pm$ 0.05 & SFG \\ 
128977 &  & 150.1122100 & 2.3140118 & 2.23 $^{+0.20}_{-0.21}$ & 23.23 $\pm$ 0.21 & SFG \\ 
129022 & ibhm52.157 & 150.0961200 & 2.3134823 & 2.05 $^{+0.12}_{-0.11}$ & 21.47 $\pm$ 0.04 & QG \\ 
132654 &  & 150.0918400 & 2.3567889 & 1.97 $^{+0.22}_{-0.21}$ & 23.04 $\pm$ 0.14 & QG \\ 
133784 &  & 150.1292900 & 2.3695824 & 2.03 $^{+0.22}_{-0.24}$ & 22.84 $\pm$ 0.13 & QG \\ 
133915 &  & 150.1335800 & 2.3703971 & 1.86 $^{+0.12}_{-0.12}$ & 21.58 $\pm$ 0.05 & SFG \\ 
134068 & ibhm46.116 & 150.1635700 & 2.3724642 & 2.02 $^{+0.10}_{-0.09}$ & 21.89 $\pm$ 0.06 & SFG \\ 
134172 &  & 150.1102600 & 2.3741243 & 2.26 $^{+0.21}_{-0.22}$ & 22.53 $\pm$ 0.12 & QG \\ 
135214 &  & 150.1841100 & 2.3863766 & 1.89 $^{+0.10}_{-0.10}$ & 22.23 $\pm$ 0.08 & SFG \\ 
135878 & ibhm53.253 & 150.1085100 & 2.3938241 & 1.94 $^{+0.09}_{-0.08}$ & 21.33 $\pm$ 0.04 & QG \\ 
139823 &  & 150.1546300 & 2.4443281 & 1.87 $^{+0.08}_{-0.08}$ & 20.69 $\pm$ 0.02 & QG \\ 
140122 & ibhm35.195 & 150.0796800 & 2.4495835 & 2.16 $^{+0.15}_{-0.17}$ & 21.88 $\pm$ 0.05 & QG \\ 
204878 &  & 150.1114000 & 2.4530017 & 2.07 $^{+0.14}_{-0.12}$ & 22.48 $\pm$ 0.10 & SFG \\ 
207084 &  & 150.1361200 & 2.4815121 & 2.05 $^{+0.14}_{-0.14}$ & 21.97 $\pm$ 0.07 & SFG \\ 
207144 &  & 150.1395000 & 2.4818728 & 1.91 $^{+0.13}_{-0.11}$ & 21.00 $\pm$ 0.04 & QG \\ 
\hline
\end{tabular}
\end{center}
\end{table*}

\section{Spectral fitting}
\label{fast}
\begin{figure*}
\centering
  \includegraphics[width=0.93\textwidth]{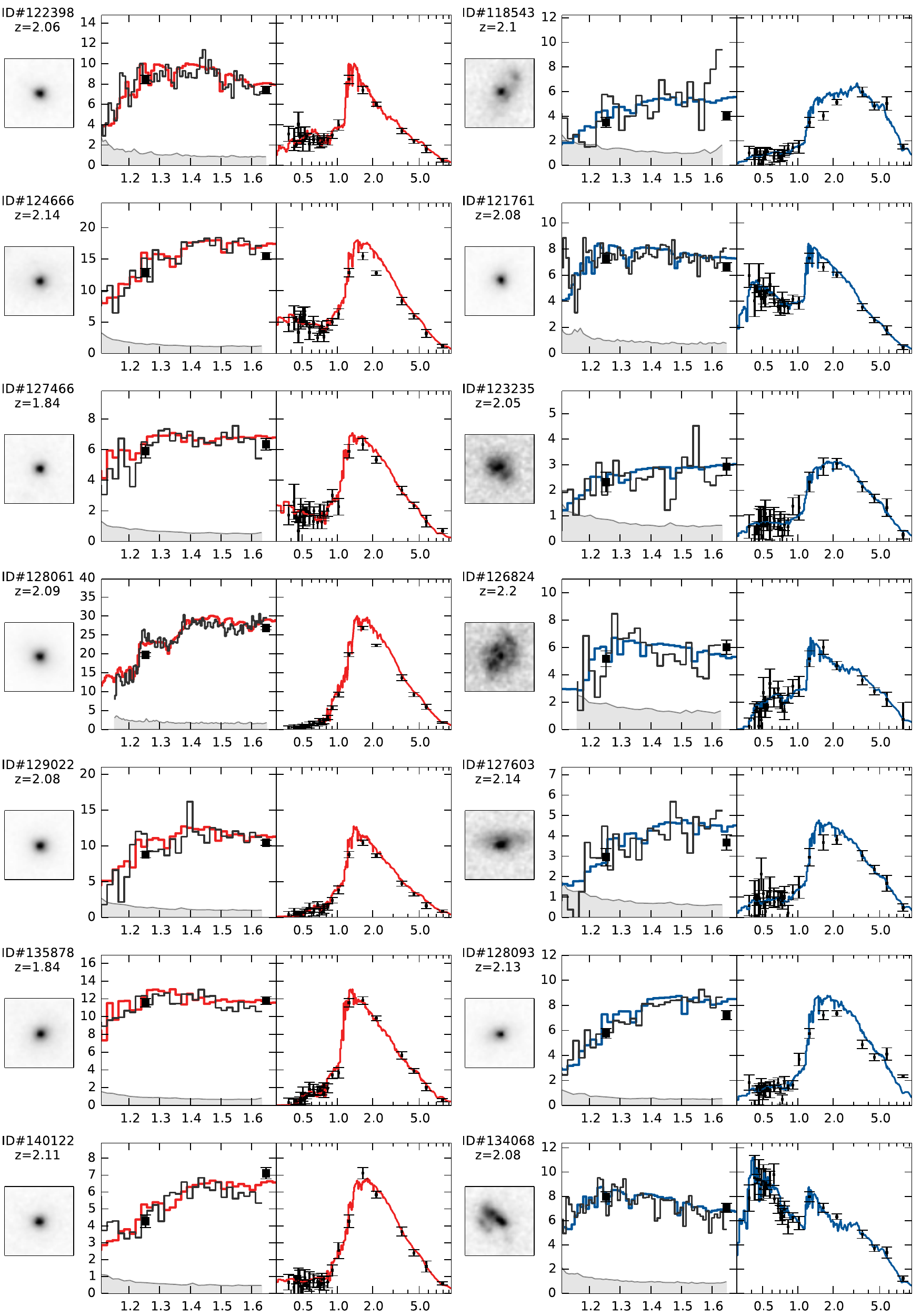}
  \caption{({\it left}) $2.4\times2.4$~arcsec$^2$ $H_{160}$-band image, ({\it middle}) 1D extracted
  grism spectrum in black and error spectrum in grey, and ({\it right}) photometric SED.
  Both the middle and right panels show wavelength in units of $\mu$m vs. $f_{\lambda}$ in units of
  $10^{-19}~\mathrm{erg}^{-1} \mathrm{s}^{-1} \mathrm{cm}^{-2} \mathrm{\AA}^{-1}$.
  The colored line shows the best-fit model from {\sc fast} convolved and rebinned
  to match the resolution of the grism spectra. Quiescent (star-forming) galaxies are shown in the left (right) column in red (blue).
   \label{fig:specSEDs}}
\end{figure*}

All galaxies in our sample were fitted by the {\sc fast} code \citep{Kriek2009a} using their photometry and photometric redshifts only. The code performs template fitting combining the photometric data with our grism spectra using exponentially declining star-formation histories (with log($\tau$) from 7.0 to 9.9 in steps of 0.1), stellar population synthesis models by \citet{BC03} and a
\citet{Chabrier2003} initial mass function.
All fits (both with and without spectra) were performed with variable metallicity among four discrete values: $Z=0.004, 0.008, 0.02, 0.05$. We used templates with a grid of ages from $7.5 < \log({\rm age / yr}) < 9.7$ in steps of 0.1, and always less than the age of the Universe at the given redshift, and with a grid of dust attenuation of $0<A_{\rm V}<2.5$ in steps of 0.1.
The subsample of sources with spectra were then fitted again with the added spectral information along with the photometry.
Before fitting the spectra, we binned them by a factor of 8. We did this to avoid being affected by differences in the resolution of the models and the actual spectra. Some of this difference was caused by morphological broadening which arises due to the fact that we are using slitless spectroscopy on extended objects. We chose to re-bin the spectra as this is much faster than having to match entire grid of templates in the fit to the resolution and sampling of the grism spectra, especially since the resolution is different for each source. The addition of spectral information improved the parameter estimation. However, the strong re-binning of the spectra made the improvement on the redshift very limited. We therefore used the spectral fits to improve the redshifts by comparing the best-fit template to the original spectrum before the re-binning. We cross-correlated the two to obtain a higher precision on the best-fit redshift. Before comparing the spectra and templates we match the resolution and sampling of the templates to those of the spectra. For spectra with low SNR we re-binned the spectra by a factor of 2 or 4 until we reached a SNR per pixel of at least 4. The uncertainty on the redshift is thus mainly determined by the spectral re-binning.
Figure~\ref{fig:specSEDs} shows the individual spectra and SEDs along with their best fitting template for the spectroscopic sample. All the parameters from the SED fits are summarized in Table~\ref{tab:fast} where we also list the strength of the 4000~{\AA} break for comparison with other studies. These are calculated using the original definition from \citet{Bruzual1983}.

\begin{figure}
  \includegraphics[width=0.50\textwidth]{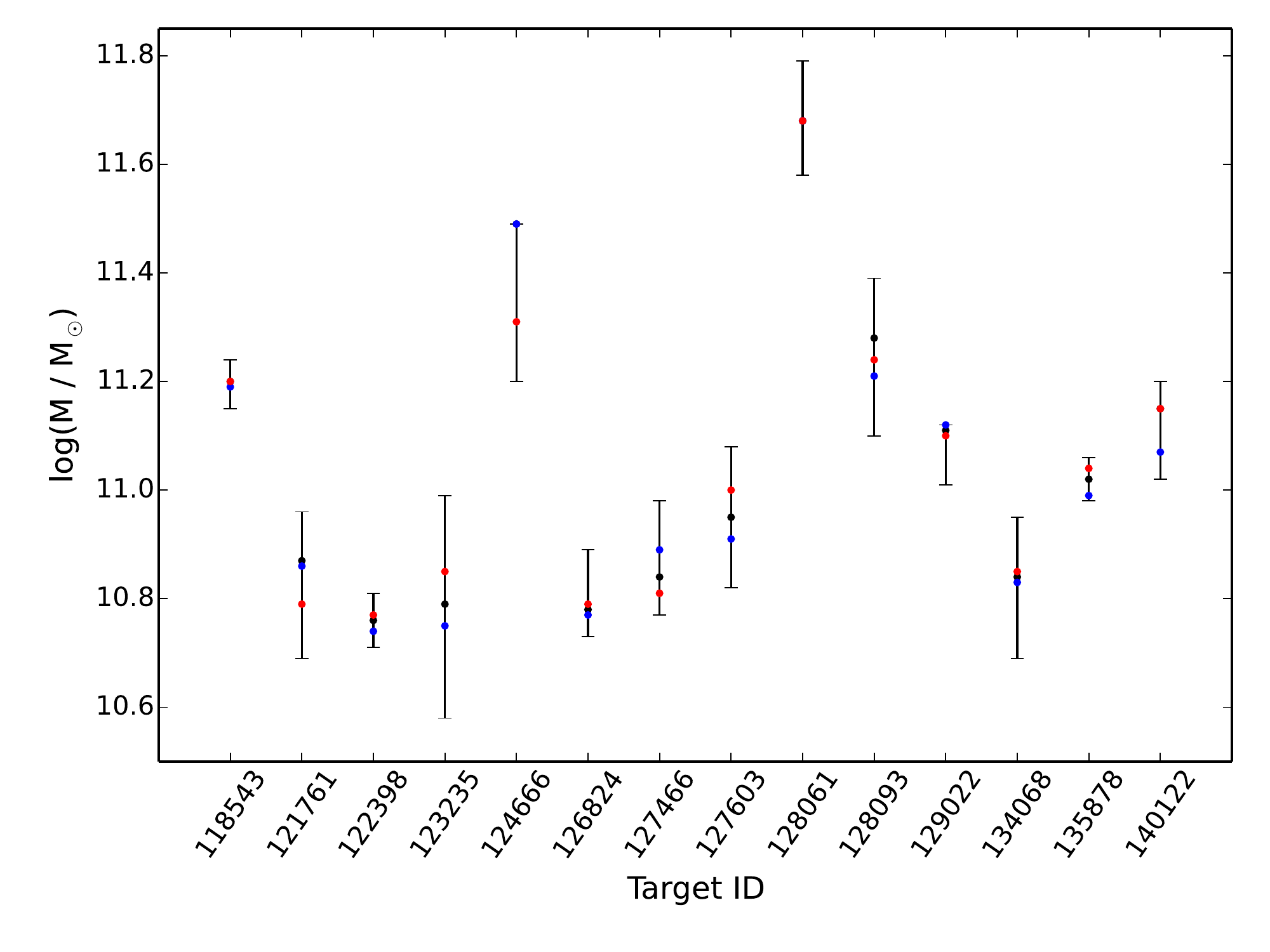}
  \caption{Effects of varying spectro-photometric redshift in the SED fits from {\sc fast}. The black points with errorbars show the best-fit stellar mass and 68\% confidence intervals with the assumed redshift for each target. The effect on stellar mass when using the upper and lower boundaries on the redshifts is shown by the red and blue points, respectively.
   \label{fig:mass_limits}}
\end{figure}

To estimate the improvement on the stellar mass with the higher precision on redshifts, we performed a last fit to the spectroscopic subsample while keeping the redshifts fixed to the spectrophotometric redshifts determined above. We note that this approach slightly underestimates the uncertainties on the masses, however, we were not able to propagate the redshift uncertainties directly using the {\sc fast} code. Instead, we performed two further fits, this time using the upper and lower $1~\sigma$ limits on the spectro-photometric redshifts. The effect of varying the redshift this way did in most cases not lead to any significant change. In Fig.~\ref{fig:mass_limits}, we show the individual mass estimates from the best-fit and the effect on the mass determination by using the upper and lower boundaries on the redshift. For all sources the variations are within the $1~\sigma$ limits from the fixed-redshift fits. We therefore neglect the effect from the redshift uncertainties on the stellar mass estimates.

In the previous section, we described the $UVJ$ method we use to select quiescent galaxies. Now with the stellar population parameters we can investigate the relation between the classification from the $UVJ$ criteria and the inferred star formation rates.
When looking at the specific star formation rate ($\rm{sSFR \equiv SFR/M_{\star}}$), we find a strong correlation between the classification of quiescent galaxies from the $UVJ$ method and galaxies with $\rm{sSFR<-10.5~yr^{-1}}$. There are only two star-forming galaxies (from $UVJ$ criteria) below this cut-off, and one quiescent galaxy above it. For the spectroscopic sample, only one star-forming galaxy lies below the sSFR limit of $-10.5~{\rm yr}^{-1}$.

\begin{figure}
  \includegraphics[width=0.48\textwidth]{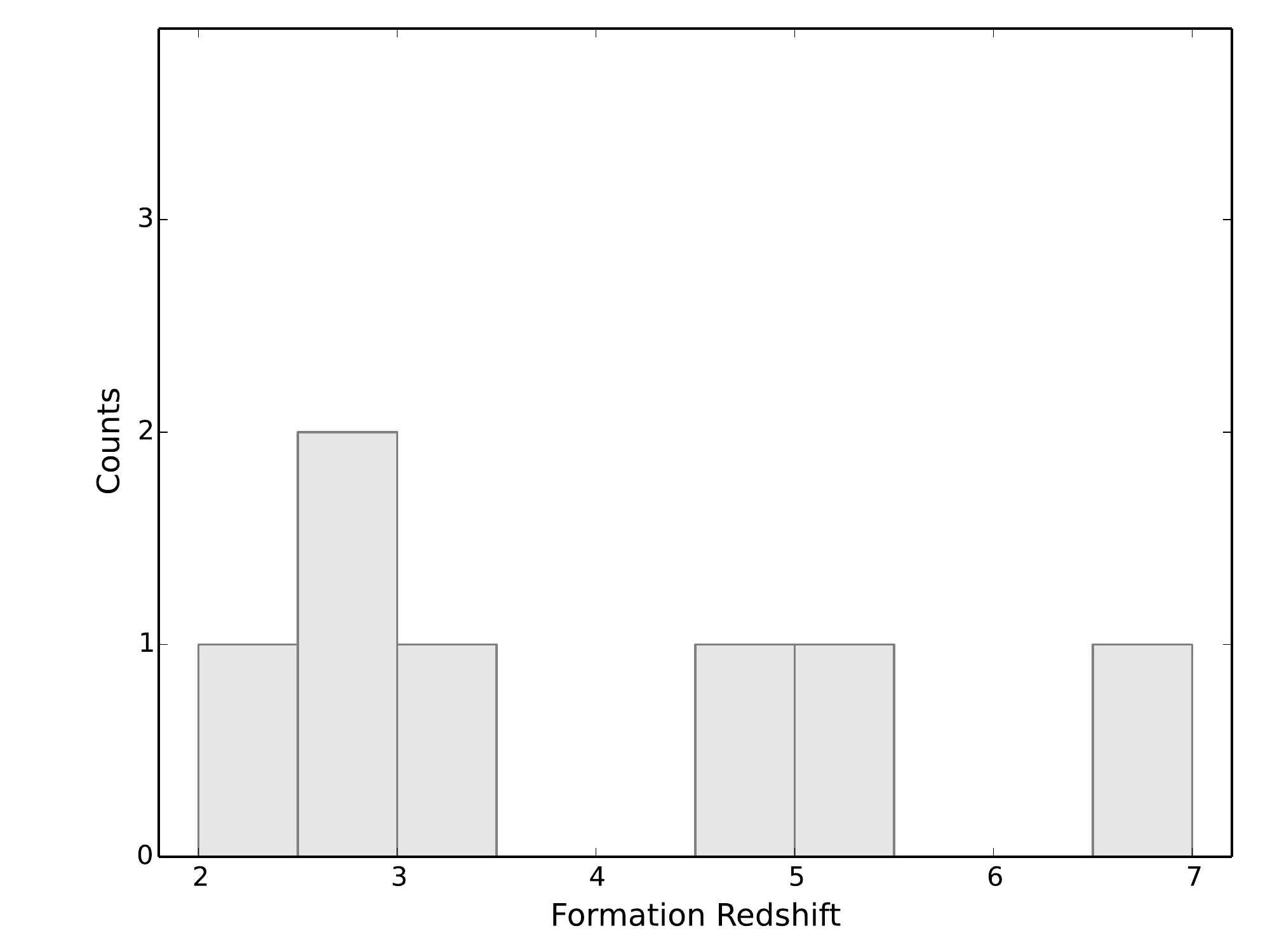}
  \caption{Distribution of formation redshifts for the quiescent galaxies. The gray distribution shows the full photometric and spectroscopic sample. The red distribution is the spectroscopic sub-sample only. Due to the much larger uncertainties on both redshift and stellar age for the photometric sample, these formation redshifts should only be considered indicative.
   \label{fig:params}}
\end{figure}

Lastly, we use the inferred ages and redshifts for the spectroscopic sample to compute formation redshifts. In Fig.~\ref{fig:params}, we show the distribution of formation redshifts for the full sample as well as for the spectroscopic sub-sample of quiescent galaxies.

\begin{table*}[h]
\caption{Parameters from SED fitting. The upper part (above the line) shows the fits with spectral data available. The redshift is in this case the spectro-photometric redshift, see text. The lower part shows the photometric sample, where $z$ here refers to the photometric redshift. \label{tab:fast}}
\begin{center}
\setstretch{1.3}
\begin{tabular}{lcccccccc}
\hline
\hline
ID & $z$ & log(M$_\star$) & log(Age) & log(Z/Z$_{\odot}$) & A$_\mathrm{v}$ & log(SSFR) & log($\tau$) & D$(4000)$ \\ 
 &  & ($M_{\odot}$) & (yr) &  & (mag) & (yr$^{-1}$) & (yr) &  \\ 
\hline
118543 & $2.10 \pm 0.05$ & $11.20^{+0.04}_{-0.05}$ & $8.80^{+0.20}_{-0.19}$ & $0.050^{+0.000}_{-0.014}$ & $1.80^{+0.20}_{-0.30}$ & $-9.72^{+0.26}_{-0.19}$ & $8.20^{+0.20}_{-0.19}$ & $3.05 \pm 1.76$ \\ 
121761 & $2.08 \pm 0.02$ & $10.87^{+0.09}_{-0.18}$ & $8.80^{+0.30}_{-0.57}$ & $0.004^{+0.015}_{-0.000}$ & $1.30^{+0.27}_{-0.43}$ & $-9.26^{+0.27}_{-0.39}$ & $8.40^{+0.36}_{-0.75}$ & $1.09 \pm 0.10$ \\ 
122398 & $2.06 \pm 0.02$ & $10.76^{+0.05}_{-0.05}$ & $8.80^{+0.19}_{-0.16}$ & $0.050^{+0.000}_{-0.040}$ & $0.20^{+0.36}_{-0.20}$ & $-10.53^{+0.09}_{-0.10}$ & $8.00^{+0.19}_{-0.16}$ & $1.30 \pm 0.13$ \\ 
123235 & $2.05 \pm 0.10$ & $10.79^{+0.20}_{-0.21}$ & $9.10^{+0.40}_{-0.78}$ & $0.004^{+0.036}_{-0.000}$ & $1.60^{+0.40}_{-0.62}$ & $-9.74^{+0.55}_{-1.02}$ & $8.60^{+0.51}_{-1.18}$ & $1.59 \pm 0.68$ \\ 
124666 & $2.14 \pm 0.05$ & $11.49^{+0.00}_{-0.29}$ & $9.40^{+0.00}_{-0.42}$ & $0.004^{+0.046}_{-0.000}$ & $0.50^{+0.21}_{-0.50}$ & $-10.62^{+0.36}_{-0.38}$ & $8.70^{+0.01}_{-0.52}$ & $1.30 \pm 0.13$ \\ 
126824 & $2.20 \pm 0.02$ & $10.78^{+0.11}_{-0.05}$ & $8.10^{+0.19}_{-0.12}$ & $0.050^{+0.000}_{-0.031}$ & $1.50^{+0.32}_{-0.20}$ & $-11.26^{+1.38}_{-1.09}$ & $7.10^{+0.38}_{-0.10}$ & $1.67 \pm 0.44$ \\ 
127466 & $1.84 \pm 0.05$ & $10.84^{+0.14}_{-0.07}$ & $9.30^{+0.20}_{-0.24}$ & $0.008^{+0.019}_{-0.004}$ & $0.50^{+0.28}_{-0.11}$ & $-10.53^{+0.18}_{-0.18}$ & $8.60^{+0.23}_{-0.24}$ & $1.34 \pm 0.35$ \\ 
127603 & $2.14 \pm 0.05$ & $10.95^{+0.13}_{-0.13}$ & $9.10^{+0.30}_{-0.61}$ & $0.020^{+0.030}_{-0.016}$ & $0.90^{+0.96}_{-0.21}$ & $-10.35^{+0.54}_{-0.50}$ & $8.40^{+0.33}_{-0.81}$ & $1.19 \pm 0.30$ \\ 
128061 & $2.09 \pm 0.01$ & $11.68^{+0.11}_{-0.10}$ & $9.40^{+0.10}_{-0.40}$ & $0.008^{+0.015}_{-0.004}$ & $0.10^{+0.42}_{-0.10}$ & $<-12.28$ & $8.20^{+0.30}_{-1.20}$ & $1.73 \pm 0.10$ \\ 
128093 & $2.13 \pm 0.05$ & $11.28^{+0.11}_{-0.18}$ & $9.00^{+0.40}_{-0.07}$ & $0.020^{+0.030}_{-0.016}$ & $1.20^{+0.15}_{-0.75}$ & $-10.26^{+0.00}_{-0.46}$ & $8.30^{+0.40}_{-0.15}$ & $1.28 \pm 0.13$ \\ 
129022 & $2.08 \pm 0.05$ & $11.11^{+0.01}_{-0.10}$ & $9.00^{+0.12}_{-0.09}$ & $0.020^{+0.012}_{-0.004}$ & $0.40^{+0.09}_{-0.40}$ & $<-12.10$ & $8.00^{+0.12}_{-1.00}$ & $1.86 \pm 0.36$ \\ 
134068 & $2.08 \pm 0.02$ & $10.84^{+0.11}_{-0.15}$ & $8.90^{+0.23}_{-0.46}$ & $0.050^{+0.000}_{-0.016}$ & $1.30^{+0.22}_{-0.17}$ & $-8.74^{+0.30}_{-0.21}$ & $9.90^{+0.00}_{-1.45}$ & $1.11 \pm 0.10$ \\ 
135878 & $1.84 \pm 0.10$ & $11.02^{+0.04}_{-0.04}$ & $8.90^{+0.06}_{-0.14}$ & $0.020^{+0.009}_{-0.007}$ & $0.70^{+0.30}_{-0.12}$ & $<-12.01$ & $7.80^{+0.17}_{-0.80}$ & $1.04 \pm 0.15$ \\ 
140122 & $2.11 \pm 0.10$ & $11.15^{+0.05}_{-0.13}$ & $9.50^{+0.00}_{-0.38}$ & $0.004^{+0.039}_{-0.000}$ & $0.50^{+0.21}_{-0.50}$ & $-11.17^{+0.27}_{-0.60}$ & $8.70^{+0.02}_{-0.36}$ & $1.40 \pm 0.19$ \\ 
 & & & & & & & & \\
\hline
 & & & & & & & & \\
119753 & 1.99 $^{+0.17}_{-0.16}$ & $11.24^{+0.17}_{-0.34}$ & $9.45^{+0.15}_{-0.62}$ & $0.008^{+0.042}_{-0.004}$ & $1.00^{+1.27}_{-1.00}$ & $-10.44^{+1.16}_{-0.78}$ & $8.80^{+1.10}_{-0.68}$ &  \\ 
123324 & 2.26 $^{+0.11}_{-0.12}$ & $10.62^{+0.31}_{-0.45}$ & $8.80^{+0.80}_{-1.18}$ & $0.050^{+0.000}_{-0.046}$ & $1.80^{+0.70}_{-0.53}$ & $-9.12^{+0.88}_{-0.54}$ & $8.50^{+1.40}_{-1.50}$ &  \\ 
123780 & 2.26 $^{+0.20}_{-0.20}$ & $10.80^{+0.32}_{-0.59}$ & $9.00^{+0.60}_{-1.29}$ & $0.004^{+0.046}_{-0.000}$ & $1.60^{+0.90}_{-1.46}$ & $<-7.86$ & $8.80^{+1.10}_{-1.80}$ &  \\ 
123817 & 2.13 $^{+0.13}_{-0.13}$ & $10.98^{+0.12}_{-0.33}$ & $9.50^{+0.10}_{-0.60}$ & $0.004^{+0.046}_{-0.000}$ & $1.00^{+0.64}_{-0.59}$ & $-9.64^{+0.78}_{-0.43}$ & $9.30^{+0.60}_{-0.89}$ &  \\ 
124168 & 2.04 $^{+0.08}_{-0.10}$ & $11.16^{+0.30}_{-0.32}$ & $9.55^{+0.05}_{-0.96}$ & $0.050^{+0.000}_{-0.041}$ & $2.20^{+0.30}_{-0.78}$ & $-9.52^{+0.60}_{-0.64}$ & $9.60^{+0.30}_{-1.41}$ &  \\ 
124686 & 2.11 $^{+0.17}_{-0.13}$ & $10.96^{+0.23}_{-0.27}$ & $9.20^{+0.40}_{-0.71}$ & $0.004^{+0.041}_{-0.000}$ & $0.50^{+1.52}_{-0.50}$ & $<-9.91$ & $8.30^{+0.70}_{-1.30}$ &  \\ 
126073 & 2.21 $^{+0.11}_{-0.09}$ & $10.67^{+0.35}_{-0.07}$ & $9.10^{+0.37}_{-0.42}$ & $0.008^{+0.042}_{-0.004}$ & $0.00^{+0.73}_{-0.00}$ & $<-11.32$ & $7.60^{+0.93}_{-0.60}$ &  \\ 
126301 & 2.05 $^{+0.11}_{-0.11}$ & $11.11^{+0.27}_{-0.17}$ & $9.20^{+0.40}_{-0.37}$ & $0.050^{+0.000}_{-0.045}$ & $1.20^{+0.96}_{-0.42}$ & $-10.09^{+0.71}_{-0.44}$ & $8.60^{+1.12}_{-0.35}$ &  \\ 
126952 & 2.28 $^{+0.27}_{-0.29}$ & $10.70^{+0.51}_{-0.37}$ & $8.35^{+1.25}_{-0.37}$ & $0.004^{+0.046}_{-0.000}$ & $2.50^{+0.00}_{-2.01}$ & $<-8.97$ & $7.30^{+2.60}_{-0.30}$ &  \\ 
127617 & 2.07 $^{+0.07}_{-0.07}$ & $11.14^{+0.22}_{-0.14}$ & $8.40^{+0.62}_{-0.50}$ & $0.008^{+0.030}_{-0.004}$ & $1.80^{+0.27}_{-0.47}$ & $-8.90^{+0.37}_{-3.36}$ & $8.00^{+1.90}_{-1.00}$ &  \\ 
128977 & 2.23 $^{+0.20}_{-0.21}$ & $11.03^{+0.27}_{-0.59}$ & $9.50^{+0.10}_{-1.72}$ & $0.020^{+0.030}_{-0.016}$ & $1.90^{+0.60}_{-1.31}$ & $-9.57^{+1.03}_{-2.69}$ & $9.40^{+0.50}_{-2.40}$ &  \\ 
132654 & 1.97 $^{+0.22}_{-0.21}$ & $10.81^{+0.21}_{-0.50}$ & $9.55^{+0.05}_{-1.26}$ & $0.004^{+0.046}_{-0.000}$ & $1.50^{+1.00}_{-1.50}$ & $<-9.82$ & $8.90^{+1.00}_{-1.90}$ &  \\ 
133784 & 2.03 $^{+0.22}_{-0.24}$ & $10.86^{+0.25}_{-0.45}$ & $9.25^{+0.35}_{-0.99}$ & $0.004^{+0.046}_{-0.000}$ & $1.20^{+1.30}_{-1.07}$ & $<-9.39$ & $8.40^{+0.86}_{-1.40}$ &  \\ 
133915 & 1.86 $^{+0.12}_{-0.12}$ & $10.87^{+0.18}_{-0.25}$ & $9.10^{+0.50}_{-0.79}$ & $0.008^{+0.034}_{-0.004}$ & $1.00^{+0.73}_{-0.44}$ & $-9.54^{+0.81}_{-0.34}$ & $8.70^{+1.20}_{-0.97}$ &  \\ 
134172 & 2.26 $^{+0.21}_{-0.22}$ & $11.13^{+0.09}_{-0.42}$ & $9.45^{+0.11}_{-0.63}$ & $0.004^{+0.046}_{-0.000}$ & $0.70^{+0.87}_{-0.70}$ & $<-10.62$ & $8.60^{+0.27}_{-1.60}$ &  \\ 
135214 & 1.89 $^{+0.10}_{-0.10}$ & $10.71^{+0.18}_{-0.17}$ & $8.85^{+0.54}_{-0.47}$ & $0.050^{+0.000}_{-0.030}$ & $1.30^{+0.37}_{-0.46}$ & $-9.22^{+0.38}_{-0.37}$ & $8.50^{+1.40}_{-0.58}$ &  \\ 
139823 & 1.87 $^{+0.08}_{-0.08}$ & $11.19^{+0.29}_{-0.06}$ & $9.20^{+0.26}_{-0.20}$ & $0.008^{+0.017}_{-0.004}$ & $0.20^{+0.49}_{-0.20}$ & $-11.50^{+0.29}_{-0.94}$ & $8.30^{+0.31}_{-0.29}$ &  \\ 
204878 & 2.07 $^{+0.14}_{-0.12}$ & $10.95^{+0.41}_{-0.24}$ & $9.15^{+0.45}_{-0.93}$ & $0.008^{+0.042}_{-0.004}$ & $2.50^{+0.00}_{-1.53}$ & $<-9.36$ & $8.70^{+1.20}_{-1.70}$ &  \\ 
207084 & 2.05 $^{+0.14}_{-0.14}$ & $10.75^{+0.25}_{-0.27}$ & $8.90^{+0.70}_{-1.04}$ & $0.004^{+0.040}_{-0.000}$ & $1.70^{+0.45}_{-0.92}$ & $-8.95^{+0.60}_{-3.31}$ & $8.90^{+1.00}_{-1.90}$ &  \\ 
207144 & 1.91 $^{+0.13}_{-0.11}$ & $11.62^{+0.07}_{-0.32}$ & $9.50^{+0.09}_{-0.51}$ & $0.020^{+0.030}_{-0.016}$ & $0.10^{+1.37}_{-0.10}$ & $-11.78^{+1.07}_{-0.16}$ & $8.60^{+0.28}_{-0.55}$ &  \\ 
\hline
\end{tabular}
\end{center}
\end{table*}

\section{Structural Fitting}
\label{galfit}
When possible we use the tabulated values from \citet{vanderWel2012}. In cases of bad fits (flag values 1, 2 and 3 in their table), we redid the fits ourselves in order to improve the fit. For this purpose, we used the same point spread functions as were used by \citeauthor{vanderWel2012} (private communication). See below for individual comments on the sources where a standard S\'ersic fit was not adequate. All neighbouring objects were included in the modelling as either individual S\'ersic components or point sources. In three cases (\#124666, \#128093, and \#207144), the fits did not converge immediately and we therefore kept the magnitude fixed to the value obtained from the CANDELS catalog. This way we were able to get the fit to converge. 

For the target \#127466, we were not able to get a good fit with a S\'ersic component due to the source being very compact. The target is furthermore surrounded by a few companions and a very spatially extended low surface brightness component. This makes it hard to separate the various components and estimate the actual background, which caused the divergent behaviour of GALFIT. Instead we tried to estimate the size using a curve-of-growth approach. We use circular apertures to infer the effective radius since the source has very low ellipticity (minor-to-major axis ratio is $0.98$). This results in a size estimate of $r_e=0.22\pm0.01$ arcsec.

Finally, we note that the target \#133915 resides in a very dense area with many compact neighbours, the size may therefore be slightly overestimated.

All the inferred sizes are included in Table~\ref{galfit} as circularized half-light radii ($r_e$) in kpc. Throughout the rest of the paper, circularized radii will be used. We use the minor-to-major axis ratio, $q$, from the GALFIT analysis to calculate the circularized radii: $r_{\rm e} = a_{\rm e}\sqrt{q}$. The details (models and residuals) of the analysis are shown in appendix \ref{galfit_plots}.

\begin{table}
\caption{Morphological Parameters from {\small GALFIT}. The top part of the table (above the horizontal line)
shows the spectroscopic sample. Below the line is the rest of the photometric sample. \label{galfit}}
\begin{center}
\begin{tabular}{lcccc}
\hline
\hline
ID & $r_e$ & S\'ersic $n$ & b/a & Ref. \\ 
 & (kpc) &  &  &  \\ 
\hline
118543 & $4.41 \pm 0.39$ & $2.57 \pm 0.30$ & $0.51 \pm 0.03$ & 2 \\ 
121761 & $1.19 \pm 0.04$ & $4.66 \pm 0.29$ & $0.87 \pm 0.02$ & 2 \\ 
122398 & $1.59 \pm 0.05$ & $6.30 \pm 0.36$ & $0.70 \pm 0.02$ & 2 \\ 
123235 & $2.78 \pm 0.11$ & $0.72 \pm 0.10$ & $0.74 \pm 0.02$ & 2 \\ 
124666$^{*}$ & $10.00 \pm 0.10$ & $9.88 \pm 0.08$ & $0.69 \pm 0.01$ & 1 \\ 
126824 & $4.39 \pm 0.13$ & $0.53 \pm 0.07$ & $0.74 \pm 0.02$ & 2 \\ 
127466 & $1.84 \pm 0.10$ & ... & $0.98 \pm 0.03$ & 1, 2 \\ 
127603 & $3.05 \pm 0.15$ & $1.37 \pm 0.13$ & $0.48 \pm 0.02$ & 2 \\ 
128061 & $2.58 \pm 0.05$ & $4.69 \pm 0.17$ & $0.82 \pm 0.01$ & 2 \\ 
128093$^{*}$ & $3.04 \pm 0.15$ & $8.53 \pm 0.98$ & $0.53 \pm 0.04$ & 1 \\ 
129022 & $1.34 \pm 0.03$ & $3.02 \pm 0.14$ & $0.87 \pm 0.02$ & 2 \\ 
134068 & $2.85 \pm 0.03$ & $0.57 \pm 0.02$ & $0.58 \pm 0.01$ & 2 \\ 
135878 & $3.42 \pm 0.20$ & $5.27 \pm 0.23$ & $0.78 \pm 0.02$ & 1 \\ 
140122 & $1.67 \pm 0.09$ & $4.92 \pm 0.50$ & $0.79 \pm 0.04$ & 2 \\ 
 & & & & \\
\hline
 & & & & \\
119753 & $2.57 \pm 0.17$ & $0.99 \pm 0.18$ & $0.36 \pm 0.03$ & 2 \\ 
123324 & $2.16 \pm 0.34$ & $3.61 \pm 0.63$ & $0.39 \pm 0.04$ & 1 \\ 
123780 & $1.73 \pm 0.17$ & $1.41 \pm 0.35$ & $0.55 \pm 0.06$ & 2 \\ 
123817 & $3.10 \pm 0.16$ & $4.61 \pm 0.19$ & $0.85 \pm 0.02$ & 1 \\ 
124168 & $3.33 \pm 0.55$ & $2.46 \pm 0.58$ & $0.68 \pm 0.05$ & 2 \\ 
124686 & $1.51 \pm 0.05$ & $5.65 \pm 0.21$ & $0.74 \pm 0.01$ & 1 \\ 
126073 & $1.58 \pm 0.09$ & $5.43 \pm 0.36$ & $0.86 \pm 0.02$ & 1 \\ 
126301 & $5.67 \pm 0.46$ & $4.18 \pm 0.20$ & $0.81 \pm 0.02$ & 1 \\ 
126952 & $3.35 \pm 0.24$ & $0.44 \pm 0.14$ & $0.43 \pm 0.04$ & 2 \\ 
127617 & $0.46 \pm 0.01$ & $8.00 \pm 0.63$ & $0.67 \pm 0.03$ & 2 \\ 
128977 & $3.36 \pm 0.14$ & $0.50 \pm 0.08$ & $0.28 \pm 0.02$ & 2 \\ 
132654 & $2.49 \pm 0.13$ & $1.46 \pm 0.15$ & $0.42 \pm 0.02$ & 2 \\ 
133784 & $2.12 \pm 0.08$ & $0.77 \pm 0.09$ & $0.51 \pm 0.02$ & 2 \\ 
133915 & $8.75 \pm 0.46$ & $5.60 \pm 0.14$ & $0.70 \pm 0.01$ & 1 \\ 
134172 & $2.54 \pm 0.40$ & $5.11 \pm 1.07$ & $0.84 \pm 0.06$ & 2 \\ 
135214 & $2.76 \pm 0.05$ & $0.64 \pm 0.04$ & $0.39 \pm 0.01$ & 2 \\ 
139823 & $1.37 \pm 0.02$ & $4.65 \pm 0.15$ & $0.44 \pm 0.01$ & 2 \\ 
204878 & $3.33 \pm 0.32$ & $1.70 \pm 0.26$ & $0.46 \pm 0.04$ & 2 \\ 
207084 & $1.25 \pm 0.04$ & $3.08 \pm 0.19$ & $0.53 \pm 0.02$ & 2 \\ 
207144$^{*}$ & $2.88 \pm 0.06$ & $2.25 \pm 0.08$ & $0.77 \pm 0.02$ & 1 \\ 
\hline
\end{tabular}
\end{center}
References: $(1)$ This work; $(2)$ van der Wel et al. (2012).\\ 
$^{*}$ Sources where the magnitude was kept fixed, see text.\\ 
\end{table}

\section{The Mass--Size Relation}
\label{mass-size}
We have used our sample of spectroscopically confirmed galaxies at redshift $z\approx2$ to investigate
the mass--size relation of quiescent galaxies at high redshift. We parameterize the relation
following Newman et al. (2012)\nocite{Newman2012} and others:

\begin{equation}
r_{e} = \gamma\left(\frac{\rm M_{\star}}{10^{11} {\rm M}_{\odot}}\right)^{\beta}
 = \gamma\, {\rm M}_{11}^{\beta}~.
\end{equation}

We fit the relation in log-log space using the errors on both the stellar masses and sizes, hence we wish to minimize the perpendicular distance from the line to each point and not simply the vertical distance. This can be expressed in terms of the likelihood estimator, assuming Gaussian uncertainties:
\begin{equation}
\ln(L) = -\sum \frac{\Delta^2_i}{2(\sigma^2_i + V)} - \frac{1}{2}\sum (\sigma^2_i + V)\, ,
\end{equation}

where $$ \Delta_i = (y_i - \beta x_i - b)/\sqrt{1+\beta^2} ~{\rm and}~ \sigma^2_i = \sigma^2_{y_i} + \beta^2\sigma^2_{x_i}~~.$$

$\Delta_i$ here denotes the orthogonal distance from the datapoint ($x_i, y_i$) to the line, and $\sigma^2_i$ denotes the projected variance for each datapoint taking into account the errors on stellar mass and size. In this notation $x$ refers to $\log(M_{\star}/10^{11}M_{\odot})$ and $y$ refers to $\log(r_e / {\rm kpc})$. The $V$ that enters in the likelihood refers to the intrinsic (Gaussian) variance perpendicular to the line, which also is a free parameter in the fit.
We fit the relation to the data using a Bayesian approach to estimate the parameters and their confidence intervals. We use flat priors on all parameters: $-10<\beta<10$, $-2<b<2$, and $-10<\ln(V)<1$.

Firstly, we fit the quiescent galaxies using only the photometric data. Hereafter we include the spectroscopic data to investigate the improvement. All the fit values are summarized in Table~\ref{tab:mass-size}. The scatter we infer from the fit is the scatter perpendicular to the line, however, the quantity we are interested in is the scatter in the sizes. We thus transform the scatter to the vertical scatter as follows: $\sigma_{\rm log\,r_e}^2 = V(1+\beta^2)$.

\begin{table}
\caption{Parameter estimates for mass--size relation of quiescent galaxies at $z\sim2$ with and without spectral data.\label{tab:mass-size}}
\begin{center}
\setstretch{1.5}
\begin{tabular}{lccc}
\hline
\hline
  & $\beta$ & $\log(\gamma/{\rm kpc})$ & $\sigma_{\rm log\,r_e}$ \\
\hline
Photometry & $0.82 \pm 0.22$  & $0.30^{+0.07}_{-0.08}$ & $0.05^{+0.09}_{-0.04}$\\ 
Phot. + spectra & $0.53^{+0.29}_{-0.21}$  & $0.29 \pm 0.07$ & $0.17^{+0.05}_{-0.04}$\\ 
\hline
\end{tabular}
\end{center}
\end{table}

The addition of spectroscopic information significantly improves the estimate of the scatter since the large uncertainty on the photometric masses leads to an underestimation of the scatter. The mass-normalized size ($\gamma$) remains almost unaffected. In the rest of the paper, we will refer to the fit obtained using the spectroscopic data combined with the full photometric sample of quiescent galaxies.
The slope is poorly constrained due to the low number of data points and the large uncertainty on the stellar masses, thus still consistent with what is found by \cite{Newman2012} at similar redshifts. Recently, \citet{vanderWel2014} studied the evolution of size, slope and scatter from redshift $z\sim2$, however, they use the effective semi-major axis and not circularized radii as in previous studies. Therefore, their results on slope and scatter cannot be directly compared to the parameters obtained in this analysis.

\begin{figure}
  \includegraphics[width=0.49\textwidth]{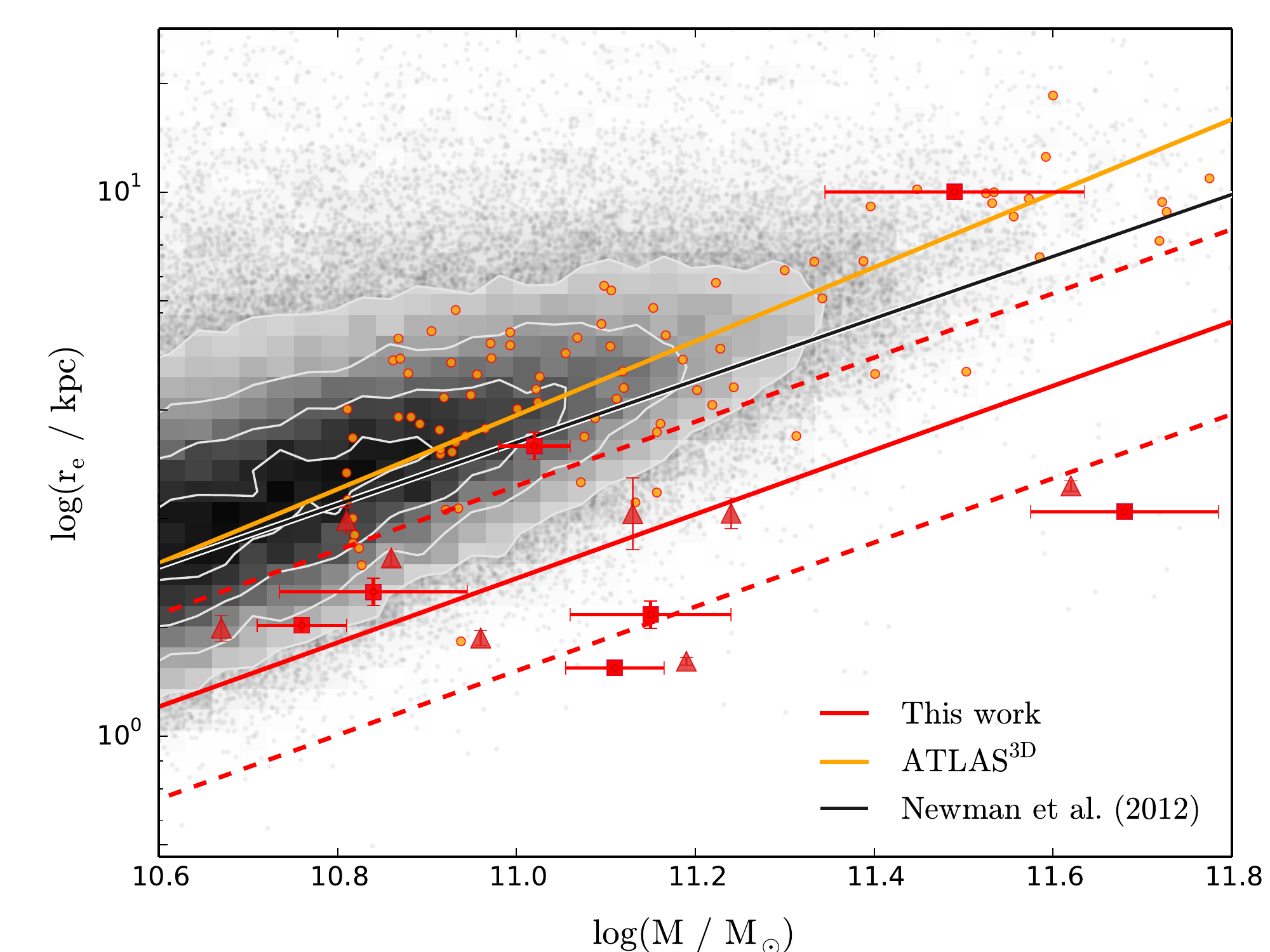}
  \caption{Mass--size relation using circularized effective radii for quiescent galaxies. The galaxies at high redshift (red points and red line) are smaller than the local relation inferred from ATLAS$^{\rm{3D}}$ data (orange points and solid line). For comparison, we show the local relation from \citet{Newman2012} as the black line. Local data from SDSS quiescent galaxies with S\'ersic indices larger than $n>2.5$ (following Shen et al. (2003)) are shown as the underlying distribution in gray. The red squares with error bars on both mass and size indicate the spectroscopic sample. The sample with only photometry available is shown as red triangles. Here we do not show the uncertainties on the mass estimates, but they are typically of the order $\sim0.2-0.3$~dex, the uncertainties are listed in Table~\ref{tab:fast} for reference.
    \label{fig:sizemass}}
\end{figure}

In Fig.~\ref{fig:sizemass}, we show the mass--size relation for the sample of quiescent galaxies with and without spectra as red squares and triangles, respectively. We compare our data to local early type galaxies with kinematical data from ATLAS$^{\rm{3D}}$
\citep[][]{Cappellari2011}. In order to compare our high redshift sample to the
that of the ATLAS$^{\rm{3D}}$-team we fit the power-law relation given above to their data
using the same mass-limit as for our data.
From the best fit to the ATLAS$^{\rm 3D}$ sample, we find the following slope
of $\beta_{0}=0.68\pm0.06$, a mass-normalized size of $\log(\gamma_{0}/{\rm kpc})=0.59\pm0.03$,
and a scatter of $\sigma_{0}=0.13\pm0.01$. For this analysis, we have used the tabulated values
from \cite{Cappellari2013a}. Specifically, we note that we used the $\log(r_{1/2})$ to infer the sizes.
The inferred slope for the local relation is significantly steeper than the value found by \cite{Shen2003} ($\beta_{\rm z=0}=0.56$). \cite{Guo2009} find a slightly higher value for the slope of $\beta_{\rm z=0}=0.70\pm0.05$ which is in very good agreement with the result from the ATLAS$^{\rm 3D}$ data.

It is clearly visible that the quiescent galaxies from this work are smaller than local quiescent galaxies at the same
mass. Moreover, the figure shows that the various samples of local galaxies infer slightly different
normalizations and slopes of the relation. 
The slope derived from the ATLAS$^{\rm{3D}}$ data is in very good agreement
with the slope derived by Guo et al. (2009). On the other hand, the relation derived
by Newman et al. (2012) and Shen et al. (2003) is much shallower. These differences are most likely caused by the different fitting techniques as well as the different approaches used to separate quiescent and star-forming galaxies.
The scatter inferred for the relation locally is found to be $\sigma_0 = 0.16$ by \citet{Newman2012} using SDSS data,
this is in good agreement with, though slightly larger than, the scatter inferred from the ATLAS$^{\rm 3D}$ data.
Given the value we infer at high redshift, $\sigma=0.17^{+0.05}_{-0.04}$, this indicates that the scatter remains
constant with redshift.

Although we cannot compare our values directly with the \citet{vanderWel2014}
study, we can still compare to their evolutionary trends. They find no evidence for evolution in scatter and slope with time, consistent with the results presented here.

\subsection{Size Evolution}
\label{size-evol}
We now take a closer look at the offset towards smaller sizes visible in Fig.\ref{fig:sizemass}
between our sample at high redshift and the local sample. This offset has been studied in great detail
\citep[e.g.,][]{Daddi2005, Trujillo2006, Toft2007, Zirm2007, Buitrago2008, vanDokkum2008,
Damjanov2011, Bruce2012, Newman2012}
and various explanations have been put forward to explain the required evolution in sizes, e.g.,
merging or feedback from quasars \citep*{Fan2008}. We here investigate a simple scenario in which
the individual galaxies themselves do not need to increase significantly in size, but rather that
the average of the population as a whole increases \citep[e.g.,][]{Cassata2011,Trujillo2012,Carollo2013,Poggianti2013}.
We use our measurements of sizes and scatter at high redshift in combinations with those from
\citet{Newman2012} to motivate the initial values for the size evolution.

\citet{Newman2012} study the size evolution of massive galaxies both star-forming and quiescent
and find that the star-forming galaxies on average are a factor of 2 larger than the quiescent population
at all times above redshift $z>0.5$. This is consistent with the fact that the star-forming
galaxies in our sample on average are a factor of $1.8$ larger than the quiescent galaxies
at a fixed mass ($10^{11}~{\rm M}_{\odot}$).
The evolution of the mean size of quiescent galaxies might then simply be driven
by the addition of larger, newly quenched galaxies at lower redshifts to the already quenched population.
\cite{Carollo2013} recently showed that the evolution of the sizes of passively evolving galaxies at $z<1$
is driven by this "dilution" of the compact population.
In order to test this picture and evaluate the effect on the scatter in sizes, we have taken the
measured sizes normalized to a stellar mass of
$10^{11}~{\rm M}_{\odot}$ from \cite{Newman2012} at redshift
$2.0<z<2.5$ and generated an initial population of quiescent (QG) and star-forming (SFG) galaxies
taking into account the observed number densities at that redshift for galaxies with comparable
masses from \citet{Brammer2011}. We have shifted the data from Newman et al. from a Salpeter
IMF to the assumed Chabrier IMF in this work. The distribution of sizes for the populations are drawn from
a log-normal distribution with an average size initially dictated by the observations for QGs while for
SFGs we simply use the fact that star-forming galaxies on average are twice as big. The distribution
of QGs is assumed to have a scatter of 0.17 dex initially, motivated by the findings in this work.
The study by van der Wel et al. (2014) show that the scatter of star-forming galaxies remains constant
with redshift and that the scatter is roughly $0.5$~dex larger than that of quiescent galaxies. We therefore
use a value of $0.22$ for the scatter of SFGs in our model. We note that the chosen value for the initial
scatter of SFGs does not change the conclusions of the model, it only enhances the increase in the modelled
evolution of the scatter.

We then simply assume that the SFGs at the given redshift will be quenched after a fixed time,
$t_{\rm quench}$, and add them to the already existing population of quiescent galaxies. For each time-step,
we generate a new population of SFGs with a mean size that is twice as big as the mean size of
the quiescent galaxies already in place, and after another $t_{\rm quench}$ these will be
added to the quiescent population. The generated number of SFGs varies according to the
observed number density of SFGs. We have assumed that the scatter of the SFG population is constant
with time and that no galaxy-galaxy interactions occur, i.e., no new massive galaxies form by merging
of lower-mass galaxies. Furthermore, we assume that galaxies maintain their sizes after they have
been quenched and that no further star formation occurs once the galaxies have been quenched.
We run this model three times for various quenching time-scales, $t_{\rm quench}$: 1.0, 1.2, and 1.5 Gyr. These time scales are consistent with the results of \citet{Bedregal2013}.

\begin{figure}
  \includegraphics[width=0.48\textwidth]{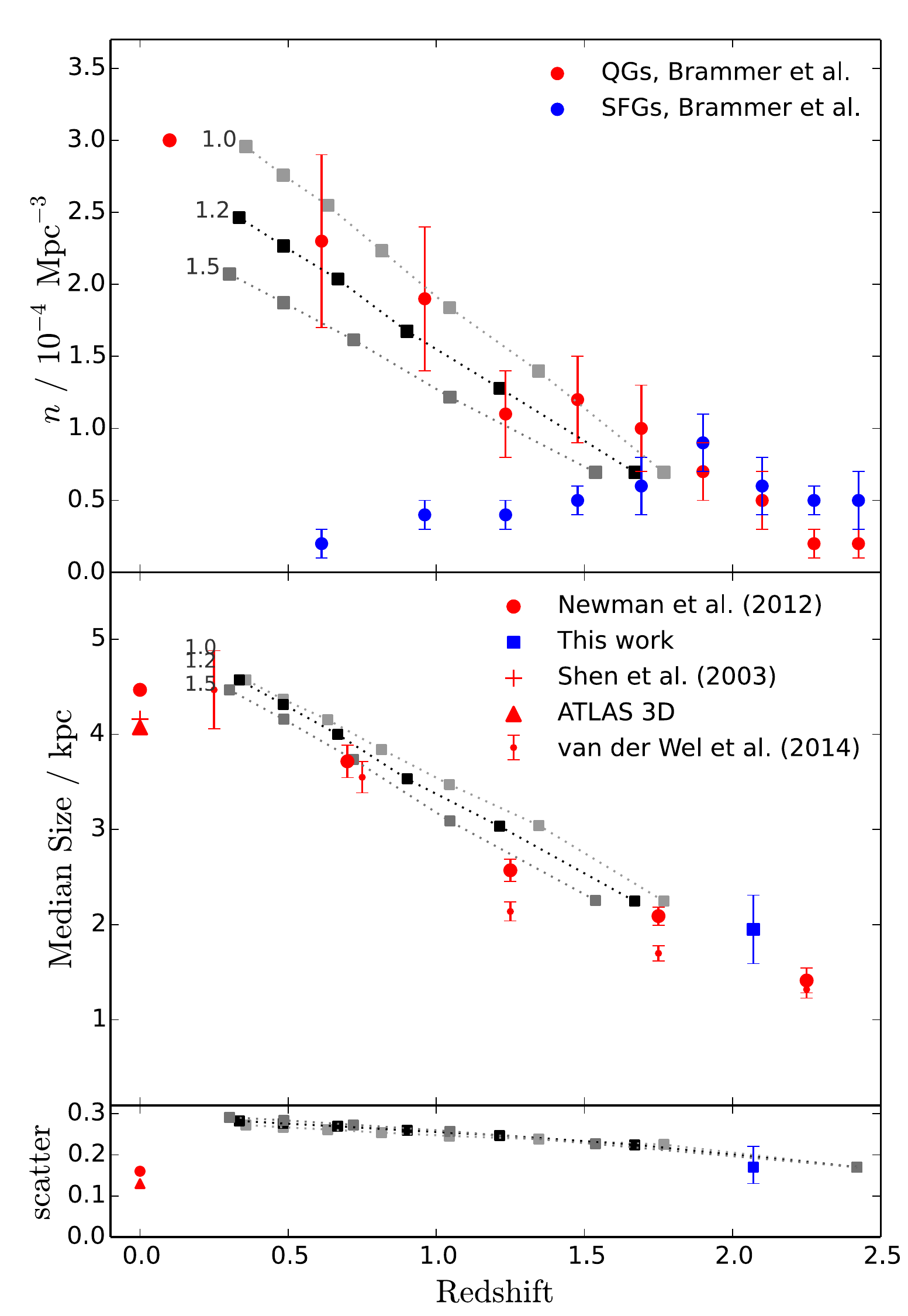}
  \caption{({\it Top}) Number density evolution with redshift. The red and blue points show
  the observed number densities for quiescent and star-forming galaxies, respectively,
  with masses $\log({\rm M / M_{\odot}})>11$ from \citet{Brammer2011}. The black and
  grey, connected points indicate our modeled evolution with varying quenching time indicated in Gyr by
  the small number at each line.
  ({\it Middle}) Average size of the quiescent galaxy population at fixed mass of $10^{11} {\rm M_{\odot}}$
  as function of redshift. The black and grey points are the same as in the top plot. The various points show
  other literature data.
  ({\it Bottom}) Modelled scatter as a function of redshift relative to the initial scatter of
  0.20 dex at redshift $z=2.4$, the first redshift-bin from Brammer et al. (2011).
  The evolution in number density and size is very well matched by the model. However, the 
  scatter of the population increases with time contrary to what is observed.
    \label{fig:sizeevol}}
\end{figure}

The results of this simple model are shown in Fig.~\ref{fig:sizeevol}. The top panel shows the evolution
in number density. The red and blue points are data from \cite{Brammer2011} for quiescent and
star-forming galaxies, respectively. The black and grey points show the modeled evolution in the number density
assuming different quenching times indicated in Gyr by the number at each of the tracks. The middle panel shows
the evolution in average size of the sample of quiescent galaxies. Data from Newman et al. (2012) is shown
in red circles, our sample is indicated by the blue square, and the local size measurements from Shen et al. (2003)
and ATLAS$^{\rm{3D}}$ are shown by the red plus and triangle, respectively. Again, the black and grey points
indicate the modeled evolution at different quenching times.
The bottom panel shows the evolution of the scatter, $\sigma$, of the distribution of quiescent galaxies
relative to the initial value at $z=2.45$.

From these assumptions, we are able to reproduce the observed increase in number density and size
of quiescent galaxies. However, the modelled scatter {\it increases} by $\sim0.15$ dex on average towards lower redshifts in contrast
with the constant scatter observed in this work and by \citet{vanderWel2014}.

\section{Discussion}
\label{discussion}

The evolution of galaxies in the mass--size plane is undoubtedly influenced 
by merging, star-formation and its cessation. As we increase the samples of 
well-studied, spectroscopically confirmed galaxies over a range of redshifts
we can forge new diagnostic tools to address the weight with which each of
these processes influences the evolution of galaxies. 

In Sect.~\ref{mass-size}, we investigated the relation between stellar mass and half-light
radius by parametrizing the relationship with a power-law. From the best fit to our quiescent
grism sample, we found the slope, $\beta = 0.53$, and the scatter, $\sigma_{\rm log\,r_{e}} = 0.17$~dex,
consistent with their $z=0$ values. From the ATLAS$^{\rm 3D}$ data and from a large SDSS
sample from the work of \cite{Shen2003}, \citet{Guo2009} and \cite{Newman2012},
a local slope and scatter of $\beta_0=0.56-0.70$ and $\sigma_0 = 0.13-0.16$~dex was inferred.
One complication in comparing samples of galaxies at different redshifts, and from different samples,
lies in the fact that the distinction between star-forming and quiescent galaxies becomes
less clear at higher redshifts. Various studies use different criteria to define quiescence, e.g.,
a cut in morphology, sSFR or rest-frame colors, which makes comparison between different datasets
non-trivial. Even at low redshift, the classification of early type galaxies is performed in different ways.

The importance of a clean separation and definition of star-forming and quiescent galaxies
becomes clear when we look at the scatter as a tool to unravel the evolution in the mass--size
relation, since the scatter is highly sensitive to outliers. Newman et al. (2012) find a scatter of
$\sigma_{\rm log\,r_{e}} = 0.26$~dex for galaxies at redshifts $2.0<z<2.5$,
much higher than what we find in our data. The large scatter observed in the Newman et al.
sample may be due, at least partly, to the uncertainty in photometric redshifts and contamination
from star-forming galaxies.\\

\subsection{Mechanisms for size growth}
\label{size-growth}

In large photometric samples it has also been shown that the slope of the
mass--size relation evolves very little from $z \sim 2$ to $z \sim 0.2-0.4$ despite there
being strong redshift evolution of the galaxy distribution in the mass--size plane
\citep[primarily a shift to larger sizes, see][\citet{McLure2013}, and \citet{vanderWel2014}]{Newman2012}.
While the unchanging slope may be theoretically plausible as the slope may reflect initial formation
rather than subsequent evolution \citep*{Ciotti2007}, the lack of evolution in the scatter
observed in this work and in the work of van der Wel et al. (2014) is puzzling.
The scatter about the mean mass--size relation should
evolve with redshift according to the underlying physical driver for the evolution in the
mass--size plane, i.e., merging, quenching, or further star formation.

Merging will typically move galaxies to higher masses and larger radii, with the direction and 
amplitude of the change in the mass--size plane determined by the mass ratio, orbital
parameters and gas content of the merger \citep*{Naab2009}. In gas-rich mergers,
the remnant may become more compact due to the gas falling to the center,
which leads to strong star formation activity \citep{Shankar2013}.
Star formation at later times (e.g., merger induced) will increase the mass,
alter the size depending on the location of the star formation, and will decrease the
mean age of the sample of \emph{quiescent} galaxies at subsequent redshifts.

Quenching of star-forming galaxies will conserve mass while the individual galaxy sizes may
even slightly decrease (as low-surface brightness star-forming regions fade) but is operating
on a separate galaxy population that has intrinsically larger sizes than most of the quiescent
galaxies already in place \citep*{Khochfar2006}. The addition of these quenched galaxies will
then drive the evolution of the mean size of the whole population without changing the
individual galaxies that have already been quenched. However, it is still not entirely clear what
happens to star-forming galaxies after they stop forming stars in terms of morphology and size;
star-forming galaxies show a variety of morphologies but the quiescent population is more
dominated by spheroidal morphologies.

\subsection{Evolution of the scatter in sizes}

Each of the above processes, in addition to directing the mass--size evolution, will affect the observed
scatter of the mass--size relation and its evolution in different ways.
Merging has been shown by \cite{Nipoti2012} to increase the scatter in the mass--size relation.
The authors show that size evolution within a dissipation-less ("dry") merger-only scenario leads
to significantly higher scatter than is observed at $z=0$ \citep*{Nipoti2012}. Mergers are certainly
on-going between $z=2$ and $z=0$ and Nipoti et al. conclude that
there must be a finely tuned balance between the different processes
in their merging model in order to reproduce the tight observed relation at $z=0$.
This type of fine tuning is not a general characteristic of the galaxy population(s) and
is extremely unlikely to persist in real-world systems. However, the
models by Nipoti et al. 2012 only consider dry mergers of spheroids, which given the diverse
population of galaxies at high redshift is an unrealistic scenario.
More recently, \citet{Shankar2013} study the size evolution of galaxies by running simulations
taking into account the various galaxy-galaxy interactions. They find that the scatter in sizes
remains constant at all times. However, they over predict the scatter by $\sim40$\,\%.

In the case of "dilution" of the population via quenching,
the scatter will increase due to the addition of a new population of larger, quiescent galaxies.
By using our toy model (see Sect.~\ref{size-evol}) for the quenching case, we have shown that
we are able to reproduce the observed increase in both number density and median size of quenched
galaxies as functions of redshift out to $z\sim2$, see Fig.~\ref{fig:sizeevol}.
However, our model shows that the scatter should increase by up to $\sim0.15$ dex in the redshift range,
$0.4 < z < 2.5$. This is inconsistent with the observations presented here and by van der Wel et al. (2014),
i.e., the observed scatter in sizes is consistent with being constant from redshift $z\approx2$ to $z=0$.

\cite{Carollo2013} have found similar results regarding the evolution of sizes and number density out to $z\sim1$ assuming the dilution of the quiescent population by younger and larger galaxies.
Other studies, however, have shown that
quenched galaxies with younger ages are not significantly larger than older quenched galaxies
\citep[e.g.,][]{Whitaker2012}, as would be predicted in a "dilution" scenario. The results of \citet{Whitaker2012}
are based on post-starburst galaxies split into "young" and "old" galaxies separated by colour cuts resulting
in median ages of the two samples of 1~Gyr and 2~Gyr, respectively, and do not as such span a very large
dynamic range in ages. Moreover, determinations of stellar ages are highly uncertain and depend strongly on metallicity, dust attenuation, and star-formation histories. The separation of "old" and "young" galaxies is thus not straight-forward and introduces large uncertainties in sample selection.

So far, most studies have focused on the role of merging only, especially dry minor merging, as the
driver of size evolution since this mechanism is very efficient in terms of increasing the size
of a galaxy without adding too much mass to the system \citep{McLure2013}. However,
\citet{Nipoti2009b,Nipoti2012} find that dry merging in a $\Lambda$CDM cosmology is
insufficient to explain the needed increase in size. As we show with our model for size evolution
in Sect.~\ref{size-evol}, the addition of larger, quenched galaxies means that each individual galaxy
needs to undergo less size-evolution. The combination of different galaxy-galaxy interactions,
both gas-rich and gas-poor, may then regulate the size-evolution of individual systems such that
the scatter remains constant through time. Also, individual systems must evolve at high redshift
as such compact galaxies locally are very rare \citep{Trujillo2009, Trujillo2014, vanderWel2014}
and merging of galaxies is an obvious mechanism for this evolution \citep{Damjanov2009,Taylor2010,Sande2011,Toft2012}.
A cascade of mergers is also the most likely way for galaxies to undergo morphological changes
from clumpy and in some cases disc-like at high redshift to spheroidal at low redshift
\citep{Naab2006,Ciotti2007,Wuyts2010}. Belli et al. (2014a) study the evolution of galaxies at fixed
velocity dispersion as the velocity dispersion has been shown to correlate with age, such that the older
populations locally have higher velocity dispersions \citep*{Belli2014a}. They target galaxies at redshifts $0.9<z<1.6$ with
a preselected spheroidal morphology. Assuming that the velocity dispersion does not
change during the lifetime of a galaxy the authors find that individual galaxies must evolve significantly,
and that size evolution driven by the addition of larger galaxies at later times is inconsistent with the velocity dispersion data.
The preselection on morphology in the study by Belli et al., however, may introduce a bias against larger
sizes since galaxies with high ellipticity or disk-like appearance on the sky would not be included. \citet{Stockton2014}
find very low axis ratios for a sample five galaxies at $z\sim0.5$ (all five having $b/a<0.5$). This may
indicate a disk-like or prolate nature of a large fraction of quiescent galaxies.

At higher redshifts, $z>2$, \citet{Belli2014b} use similar methods to study the evolution of galaxies at
fixed velocity dispersion \citep*{Belli2014b}. Though their sample at high redshift is not complete,
they conclude that a strong evolution of individual systems is needed
in order to reconcile the high-redshift size-distribution with the locally observed distribution. Furthermore, the inferred size evolution
happens at too high a rate compared to the expectations from dry, minor merging \citep{Oser2012, Nipoti2012}.
\citet{Newman2012} find consistent results, that is, at lower redshifts $z\la1$ the rate of size evolution
is consistent with dry, minor merging, while at higher redshifts the size increase happens too fast to be driven
purely by minor merging.

The fast increase in sizes at high redshift might be explained by the addition of larger, recently quenched galaxies
as the number of star-forming galaxies at this epoch in cosmic time is comparable or even dominating. Under the assumption that the galaxy population turns off star formation more or less simultaneously, quenching of
a big population of star-forming galaxies (in comparison to the size of the passive population in place) will cause a large increase in the median size of the distribution of quiescent galaxies, see Sect.~\ref{size-evol}. This effect will be stronger at higher redshifts when the number density of star-forming galaxies is high. At later times, $z\la1.5$, the quiescent population starts to dominate and hence the "dilution" effect contributes less. This is consistent with the scenario from Newman et al. and Belli et al., in which the size evolution of galaxies at lower redshifts is dominated by minor merging.

In order to study the evolution of galaxies and disentangle the various processes, high resolution, cosmological
simulations are needed, which take into account both gas-rich and gas-poor galaxy interactions
on the entire population of star-forming and quenched galaxies.
These may be able to reproduce the lack of evolution in the scatter.\\

\section{Conclusion}

We have presented for the first time for a {\it spectroscopic} sample that the slope and
scatter of the mass--size relation at $z=2$ are consistent with their local values. We use the fact that
the scatter remains constant from $z=2$ to $z=0$ as a tool to study the evolutionary mechanism
that drives the size-increase of this population. We find that while the addition of larger galaxies
quenched at later times can explain the increase of the average size of the population the scatter
increases in contrast with the results presented here. Other processes, such as the combined influence
from dry and wet mergers, is therefore needed to keep the scatter constant and to
make the number density of the most compact galaxies evolve in a way that is consistent with their
rarity in the local Universe.

\acknowledgements
The Dark Cosmology Centre is funded by the DNRF. JK acknowledges
support from an ESO studentship. ST and AZ acknowledge support from
the Lundbeck foundation.
This work is based on observations taken by the 3D-HST Treasury Program
(GO 12177 and 12328) with the NASA/ESA HST, which is operated by the
Association of Universities for Research in Astronomy, Inc.,
under NASA contract NAS5-26555.
This work is based on observations taken by the CANDELS Multi-Cycle
Treasury Program with the NASA/ESA {\it HST}, which is operated by
the Association of Universities for Research in Astronomy, Inc., under
NASA contract NAS5-26555.

\appendix
\section{The discarded spectra}
\label{sample:notes}
The discarded spectra can be grouped into the following categories:

\noindent (i) spectra with bad background subtraction

\#133784, \#133915, \#207144. These have breaks in the blue side of the spectra which could not be corrected for.\\

\noindent (ii) spectra that are heavily noise dominated where more than half the pixels have SNR$<1$:

\#123780, \#124168.\\

\noindent (iii) highly contaminated:

\#132654, \#126073.\\

\noindent (iv) flux-calibration issues, i.e., mismatch between the photometric $J$ and $H$ band:

\#119753, \#134172.\\

\noindent (v) targets that fall outside our sample criteria after adding the spectral data to the fits:

\#124686, \#126301, \#135214, \#204878, \#127617.\\
The target \#127617 presents a strong and broad emission line in the spectrum, which indicates that this is not a passively evolving galaxy. We therefore remove it from our sample. The others in the category fall below the redshift definition when the spectra were fitted together with the photometry.\\

\noindent For the following targets, no spectra were extracted:

\#123324, \#123817, \#126952, \#128977, \#139823, and \#207084.

\begin{figure}
\centering
  \includegraphics[width=0.85\textwidth]{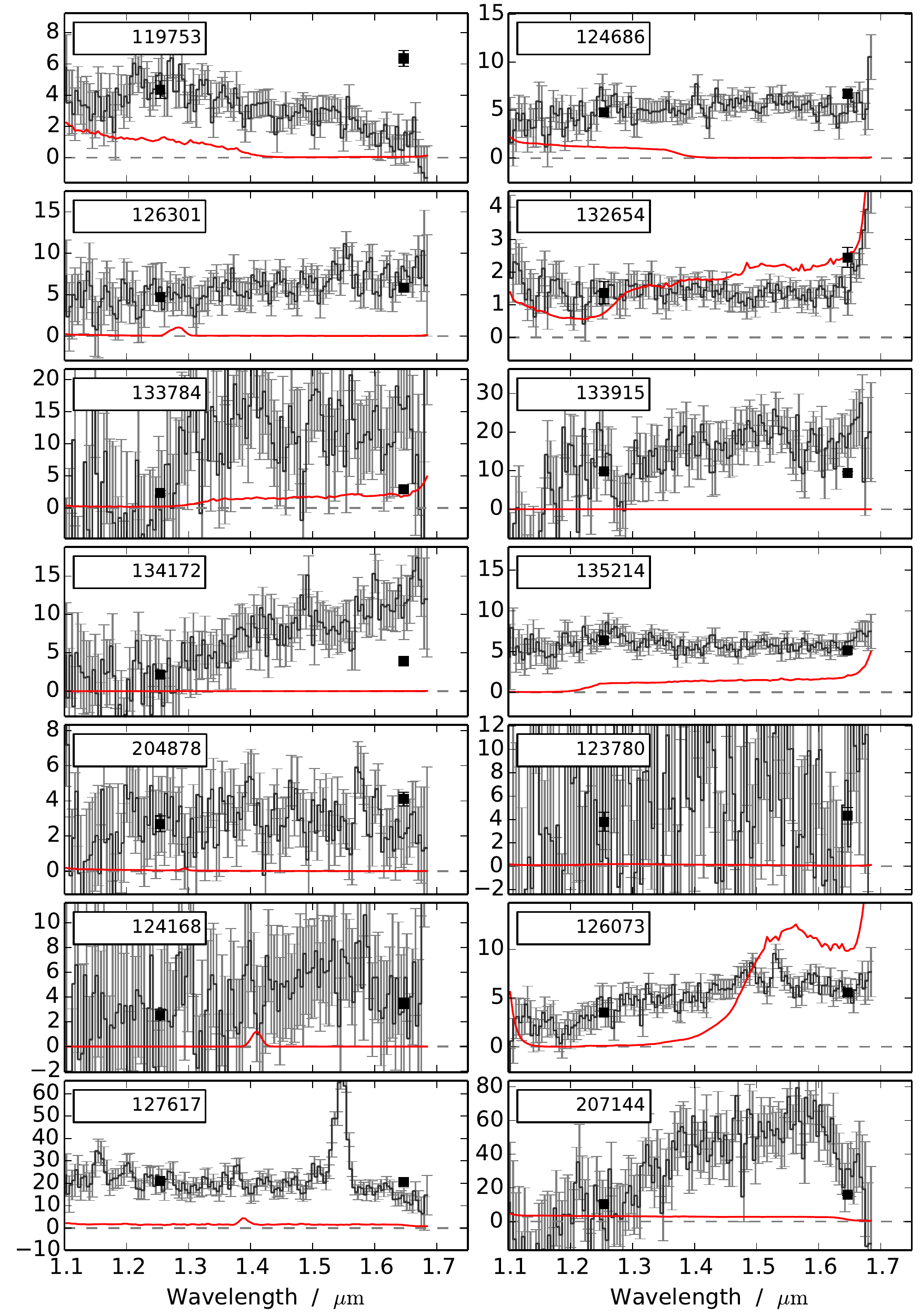}
  \caption{Discarded spectra from the sample, for individual notes see text. The spectra are shown as black with grey error bars in units of $10^{-19}~\mathrm{erg}^{-1} \mathrm{s}^{-1} \mathrm{cm}^{-2} \mathrm{\AA}^{-1}$. The red line shows the contamination estimate.
  \label{discarded_spectra}}
\end{figure}

\clearpage

\section{GALFIT results}
\label{galfit_plots}

\begin{figure*}[b]
\centering
  \includegraphics[width=0.87\textwidth]{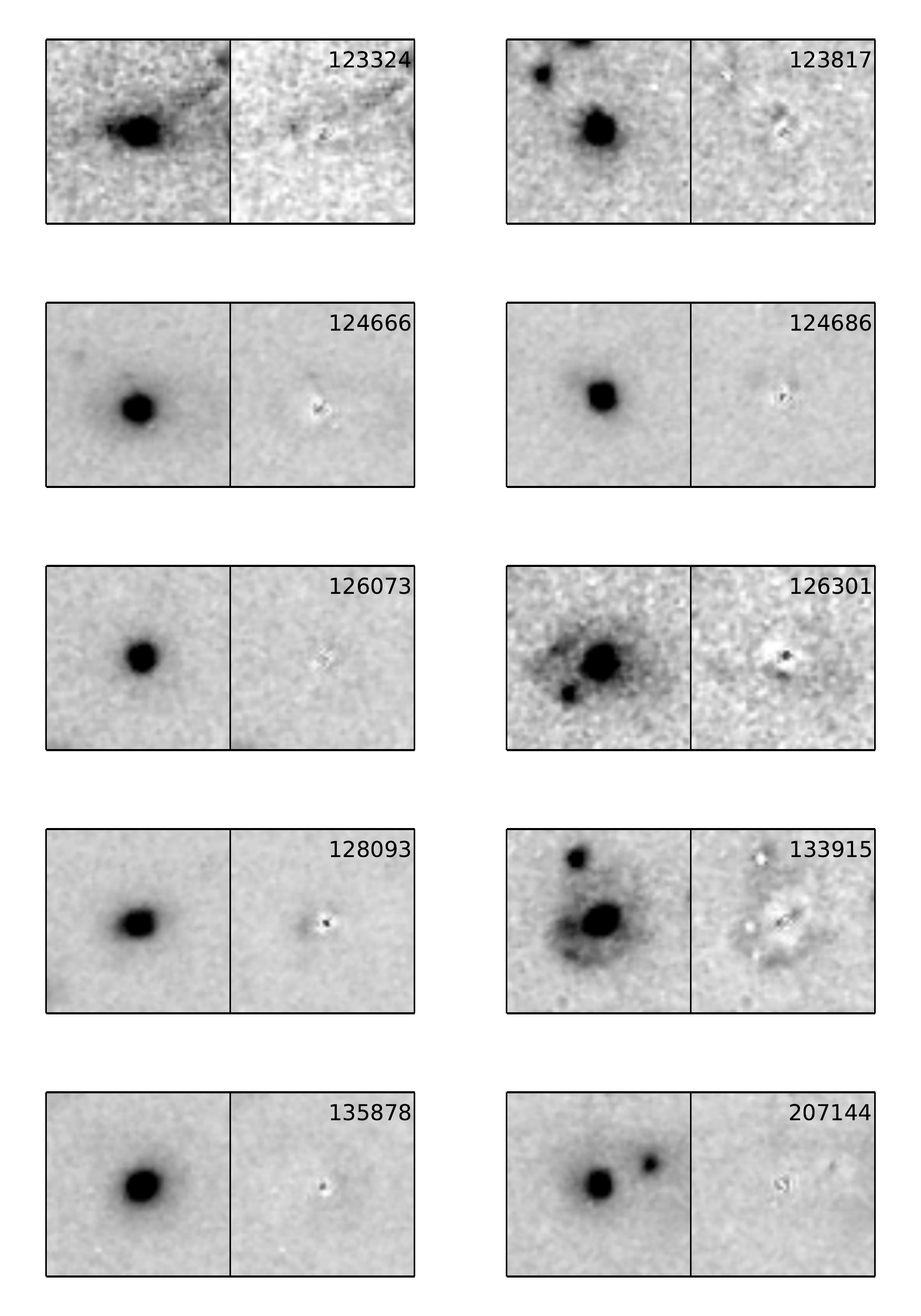}
  \caption{$H_{160}$ images from the CANDELS data ({\it left}) and the residuals after subtracting the best-fit GALFIT model ({\it right}).
  \label{galfit_residuals}}
\end{figure*}

\clearpage

\bibliographystyle{apj}

\end{document}